\documentclass[letterpaper]{JHEP3}
\usepackage{amssymb}

\usepackage{epsfig}

\usepackage{amsfonts}
\usepackage{amsmath}

 \numberwithin{equation}{section}
 
\usepackage{graphicx}

\usepackage{amssymb}
\usepackage{epsfig} 
\usepackage{amsfonts}
\usepackage{graphicx}
\usepackage{amsmath}

\newcommand{\insertplot}[5]{\begin{figure}
 \hfill\hbox to 0.05in{\vbox to #5in{\vfill
 \inputplot{#1}{#4}{#5}}\hfill}
 \hfill\vspace{-.1in}
 \caption{#2}\label{#3}
 \end{figure}}
 \newcommand{\inputplot}[3]{
 \special{ps: plotfile #1}
}
\newcounter{fig}

\newcommand{\beq}{\begin{equation}}
\newcommand{\eeq}{\end{equation}}
\newcommand{\beqs}{\begin{eqnarray}}
\newcommand{\eeqs}{\end{eqnarray}}

\numberwithin{equation}{section}
\newcommand{\be}{\begin{equation}}
\newcommand{\ee}{\end{equation}}
\newcommand{\bea}{\begin{eqnarray}}
\newcommand{\eea}{\end{eqnarray}}
\newcommand{\Ord}[2]{\mathcal O \left(#1\right)^{#2}}

\usepackage{graphicx}

\abstract{ 
We construct uniform black-string solutions in Einstein-Gauss-Bonnet gravity
for all dimensions $d$ between five and ten
and discuss their basic properties.
Closed form solutions are found by taking the
Gauss-Bonnet term as a perturbation from pure Einstein
gravity. Nonperturbative solutions are constructed by solving  numerically the equations of the model.
The Gregory-Laflamme instability of the black strings is explored via linearized perturbation theory.
Our results indicate 
that new qualitative features occur for $d=6$, in which case
 stable configurations  exist for large enough values of the Gauss-Bonnet coupling constant.
For other dimensions, the black strings are dynamically unstable and have also
a negative specific heat.
We argue that this provides an explicit realization of the Gubser-Mitra conjecture,
 which links local dynamical and thermodynamic stability. 
Non-uniform black strings in Einstein-Gauss-Bonnet theory are also constructed in six spacetime dimensions.
 
 }
\preprint{ }

\title{Einstein-Gauss-Bonnet black strings}
\author{Yves Brihaye$^{1}$\thanks{E-mail: \texttt{Yves.Brihaye@umons.ac.be}}, 
Terence Delsate$^{1}$\thanks{E-mail: \texttt{terence.delsate@umons.ac.be}}~~and\
 Eugen Radu$^2$\thanks{E-mail: \texttt{radu@theorie.physik.uni-oldenburg.de}}  \\
$^{1}$Physique-Math\'ematique, Universit\'e de Mons, Place du Parc, B-7000 Mons, Belgique\\
$^{2}$ Institut f\"ur Physik, Universit\"at Oldenburg, Postfach 2503 D-26111 Oldenburg, Germany}

\begin{document}

\section{Introduction}
In recent years it has been realized that, 
even at the classical level, 
the physics of black objects in higher dimensional spacetimes is
 much richer than
in four dimensions.
In order not to contradict observational evidence, 
the extra dimensions are usually supposed to be compactified
on small scales. 
In such $d$-dimensional manifolds with $n$ compact dimensions,
black holes can either be localized in the compact
dimensions, or the black hole horizon can wrap the
compact dimensions completely \cite{Obers:2008pj}. The static black holes which are
localized in the compact dimensions have the horizon
topology of a $(d-2)$-sphere, $S^{d-2}$, and are usually called �caged�
black holes.
 For small values of the mass, these solutions look like the Schwarzschild black hole in $d$ dimensions.
In contrast, when the horizon wraps
the compact dimensions, the horizon topology reflects the
topology of the compact manifold. In the simplest $n=1$ case,
the single compact dimension is simply a circle, $S^1$. Then the second type of
black holes have the horizon topology of a torus,
$S^{d-3} \times S^1$, and are referred to as black strings.

In vacuum Einstein gravity, a uniform black string 
(UBS) is found by adding a trivial direction to a vacuum
black hole in $d-1$ dimensions. 
A static black string exists for all values of the mass, 
although it is unstable below a critical value,
as shown by Gregory and Laflamme (GL) \cite{Gregory:1993vy}.
Following this discovery, a branch of non-uniform black string (NUBS)
solutions (thus with a dependence on the extra dimension) 
was found perturbatively from the critical
GL string in five \cite{Gubser:2001ac}, six \cite{Wiseman:2002zc} 
and in higher, up to sixteen, dimensions \cite{Sorkin:2004qq}. 
This non-uniform branch was numerically extended into the full nonlinear regime
for $d=5$ \cite{Kleihaus:2006ee} and $d=6$ \cite{Wiseman:2002zc,Kleihaus:2006ee}.
In Ref. \cite{Sorkin:2006wp}
NUBSs were constructed for all dimensions up to eleven.

All these results concern the case of Einstein gravity theory and its 
 extension  with abelian fields and scalars \cite{Kudoh:2005hf},
 \cite{Frolov:2009jr}, \cite{Kleihaus:2009ff}.
 However, for a spacetime dimension $d>4$, the Einstein gravity presents a natural generalisation
-- the so called Lovelock theory, constructed 
from vielbein, the spin connection and their exterior derivatives without using the Hodge dual,
such that the field equations are second order  \cite{Lovelock:1971yv}, \cite{Mardones:1990qc}.
Following the Ricci scalar, the next order term in the Lovelock hierarchy is the Gauss-Bonnet (GB) one,
which contains quadratic powers of the curvature.
As discussed in the literature, this term appears as the first curvature stringy
correction to general relativity~\cite{1,Myers:1987yn}, when assuming
that the tension of a string is
large as compared to the energy scale of other variables.  
The Einstein-Gauss-Bonnet (EGB) equations contain no higher derivatives of the metric tensor than second order
and the model has proven to be free of ghost when expanding around flat space.
Inclusion of a GB term in the gravity action leads to a variety of 
new features  (see \cite{Garraffo:2008hu}, \cite{Charmousis:2008kc}
for recent reviews of the  higher order gravity theories and their solutions).

Although the generalization of the spherically symmetric Schwarzschild solution
in EGB theory has been known for quite a long time \cite{Deser}, \cite{Wheeler:1985nh}, 
the issue of  solutions with compact extra dimensions is basically unexplored.
The only case discussed in a systematic way in the literature  are the $d=5$
UBSs in \cite{Kobayashi:2004hq}
and their $p-$brane generalizations\footnote{Note that these 
configurations 
are very different from the warped EGB solutions as discussed for instance in \cite{kastor}, \cite{char}.} 
in \cite{Sahabandu:2005ma}. 
The results there show  the existence of a number of new features in this case, for example
the occurrence
of a minimal value of the black strings' mass for a given 
GB coupling constant $\alpha$.
The stability of these configurations was considered in \cite{Suranyi:2008wc}.
However, the study of (asymptotically flat-) EGB black holes reveals that the case $d=5$ is rather special
\cite{Deser}, \cite{Wheeler:1985nh} (see also Appendix A).
Moreover, the results in \cite{Brihaye:2008xu} for $d\geq 5$ UBSs with Anti-de Sitter asymptotics show that
the properties of such solutions
depend on the number of spacetime dimensions.
Therefore it is of interest to study  EGB black strings for $d>5$ as well.

The purpose of this work is to discuss the main features of the
black string solutions in EGB theory for a number of dimensions between five and ten.
Starting with configurations without a dependence on extra dimension,
we first construct closed form solutions, by treating the effects of the GB term 
as a perturbation around the Einstein gravity configurations.
The nonperturbative UBSs are found by solving numerically the EGB equations with suitable
boundary conditions. For $d=5$, the results in \cite{Kobayashi:2004hq} are recovered;
however, no mass gap appears for $d>5$ configurations.
We shall address also the question of classical and thermodynamical stability of the UBSs. 
An interesting result here is the existence of black strings stabilized by the  
GB term in $d=6$ spacetime dimensions.
 Black strings with a dependence on the extra dimension will be constructed for $d=6$ only.

 The outline of the paper is as follows:
 in the next Section we explain 
the model and present general expressions for the physical quantities. 
 The  uniform black string solutions 
are discussed in Section $3$ where we show approximate
closed form solutions and results obtained by numerical calculations.
 In Section 4 we explore via linearized perturbation theory 
the GL instability of the black string solutions.
Non-uniform black string solutions in $d=6$ EGB theory are constructed in
Section 5. We give our conclusions and remarks in the final Section.  
  The Appendix contains a discussion of the basic properties of the EGB generalizations
  of the Schwarzschild solution in $d$-dimensions, 
  which helps us to better understand the 
  properties of the black strings in $d+1$ dimensions.
  
\section{ The model}

\subsection{ The action and field equations}
We consider the $d$-dimensional Einstein-Hilbert action supplemented by the GB term:
\be
I = \frac{1}{16\pi G}\int \sqrt{-g}\left( R + \frac{\alpha}{4} L_{GB}  \right) d^d x,
\label{action}
\ee
with
\be
L_{GB}=R^2 - 4 R_{\mu \nu}R^{\mu \nu} + R_{\mu \sigma \kappa \tau}R^{\mu \sigma \kappa \tau} ,
\ee
where $R$ is the Ricci scalar, $g$ is the determinant of the metric, $R_{ab}$ 
is the Ricci tensor, $R_{\mu \sigma \kappa \tau}$ is the Riemann tensor.
The constant $\alpha$ in (\ref{action}) is the GB coefficient with dimension $(length)^2$ and is positive
in the string theory.

The variation of the action (\ref{action}) with respect to the metric
tensor results in the EGB  equations 
\begin{eqnarray}
\label{eqs}
E_{\mu \nu }=R_{\mu \nu } -\frac{1}{2}Rg_{\mu \nu}
+\frac{\alpha}{4}H_{\mu \nu}=0~,
\end{eqnarray}
where
\begin{equation}
\label{Hmn}
H_{\mu \nu}=2(R_{\mu \sigma \kappa \tau }R_{\nu }^{\phantom{\nu}%
\sigma \kappa \tau }-2R_{\mu \rho \nu \sigma }R^{\rho \sigma }-2R_{\mu
\sigma }R_{\phantom{\sigma}\nu }^{\sigma }+RR_{\mu \nu })-\frac{1}{2}%
L_{GB}g_{\mu \nu }  ~,
\end{equation}
is the Lanczos (or the Gauss-Bonnet) tensor.

A general feature of the EGB gravity is the presence of two branches of solutions, distinguished by their behaviour
as $\alpha \to 0$ \cite{Deser}. However, in this paper we shall restrict to the  branch of solutions 
which present a well defined Einstein gravity limit.

\subsection{Black strings in EGB theory: general formalism}  

The simplest solution  of EGB theory corresponds  to the generalization of the Schwarzschild black hole.
 As we shall see, in $d$-dimensions, such a configuration  shares a number of basic features with the 
 EGB black strings in $d+1$ dimensions.
The basic properties of this solution are discussed in Appendix A.
For example, it has an event horizon of $S^{d-2}$ topology
 and approaches asympotically  the Minkowski spacetime background ${\cal M}^{d}$.

However, in this work we are interested\footnote{Note that, for small mass, the Schwarzschild-GB solution is expected to provide
an approximation of the Kaluza-Klein caged black holes.}
in solutions approaching asymptotically the 
$d-1$ dimensional Minkowski-space times a circle, ${\cal M}^{d-1}\times S^1$.
The line element of this background is
\be
\label{KK-metric}
ds^2 =-dt^2 + dr^2  + dz^2   + r^2 d\Omega_{d-3}^2,
\ee
where the direction $z$ is periodic with period $L$,
 $r$ and $t$ are the radial and time coordinates, respectively, while  $d\Omega^2_{d-3}$ is the unit metric on $S^{d-3}$.

The physical quantities  of a configuration that can be measured asymptotically far away in the
transverse space are the mass  $M$  and the tension ${\mathcal T}$ in the direction of the circle.
 Similar to Einstein gravity, these quantities are defined in terms of two constants
 $c_t,c_z$  which enter the asymptotics of the metric functions
 \begin{eqnarray}
\label{as2}
&g_{tt}\simeq -1+\frac{c_t}{r^{d-4}},~~~g_{zz}\simeq 1+\frac{c_z}{r^{d-4}}.
\end{eqnarray} 
The mass and tension of a black string solution are given by\footnote{For discussions
of the computation of charges in EGB theory without a cosmological
constant, see \cite{Deser:2002rt}.}
\begin{eqnarray}
\label{2} 
M=\frac{V_{d-3}L}{16 \pi G}\left[(d-3)c_t-c_z \right],
~~{\mathcal T}=\frac{V_{d-3}}{16 \pi G}\left[c_t-(d-3)c_z \right],
\end{eqnarray}
where $V_{d-3}$ is the area of the unit $S^{d-3}$ sphere.

One can also define a relative tension $n$ 
(also called the relative binding energy) 
\begin{eqnarray}
\label{3}
n=\frac{{\mathcal T} L}{M}=\frac{c_t-(d-3)c_z}{(d-3)c_t-c_z},
\end{eqnarray}
which measures how large the tension is relative to the mass. 
Uniform string solutions in vacuum Einstein gravity have $c_z=0$ and thus a 
relative tension $n=1/(d-3)$.
However, we shall see that this quantity is no longer constant when adding a GB term.

The black string are higher dimensional black objects
possessing an event horizon.
For the solutions in this work, the  event horizon is
 located at a constant value of  the radial 
coordinate  $r=r_h$, with $g_{tt}(r_h)=0$. 
Similar to the black hole case, a black string
can be considered as a
thermodynamical system, with entropy and temperature.
The Hawking temperature of the solutions is given by
\begin{eqnarray}
\label{TH-gen} 
T_H=\frac{ \kappa}{2 \pi},
\end{eqnarray}  
with $\kappa$ the surface gravity.
The general results in \cite{Wald:1993nt} shows that the entropy of a black object
($i.e.$ also of a black string) in EGB theory can be written
as a integral over the event horizon
\begin{eqnarray}
\label{S-Noether} 
S=\frac{1}{4G}\int_{\Sigma_h} d^{d-2}x \sqrt{ h}(1+\frac{\alpha}{2}\tilde R),
\end{eqnarray} 
where $ h$ is the determinant of the induced metric on the horizon and $\tilde R$ is the event horizon curvature.

 The solutions should obey the first law of thermodynamics, which for Kaluza-Klein asymptotics
 contains an extra work term:
\be
dM = T_H dS + \mathcal T dL.
\label{firstlaw}
\ee
As we shall see, a Smarr relation relating asymptotic and 
event horizon quantities exists only for EGB
black strings without a dependence on the $z$-coordinate.

\section{ Uniform black string solution}
 In this section we study EGB black strings which are translationally invariant
 along the $z-$direction.
In Einstein gravity, such configurations are constructed
by adding a flat direction to any   vacuum black hole in $d-1$ dimensions.  
However, it is straightforward to check that this simple construction does
not work in the presence of a GB term in the 
action,
 unless the solutions are conformally flat.  
 Therefore the metric function $g_{zz}$ (and thus  the proper length along the compact dimension)
 has a dependence on the radial coordinate $r$.

 \subsection{The metric ansatz and field equations}

 The uniform solutions are constructed in Schwarzschild-like coordinates, 
 with the following metric ansatz:
\be
\label{ansatz}
ds^2 = -b(r) dt^2 + \frac{dr^2}{f(r)} + g(r) d\Omega_{d-3}^2 + a(r) dz^2,
\ee
where $a,b,f,g$ are functions of the radial coordinate $r$.

Substituting (\ref{ansatz}) into the field equations (\ref{eqs}) leads to a system of four equations for 
the functions $a,b,f,g$. 
Fixing the arbitrariness of the metric gauge by taking $g(r)=r^2$, 
these equations read
 \begin{eqnarray}
\label{eqa} 
&&a''+(\frac{a'}{2}+(d-3)\frac{a}{r})\frac{f'}{f}-\frac{a'^2}{2a}+(d-3)(d-4)\frac{a(f-1)}{r^2 f}
+(d-3)\frac{a'}{r}
\\
\nonumber
&&{~~~~~~~~~~}
-\frac{\alpha}{4}\frac{(d-3)(d-4)a}{r^2 f}
\bigg(
2(f-1)f\frac{a''}{a}+ 2(d-5)(f-1)f\frac{a'}{r a}
\\
\nonumber
&&{~~~~~~~~~~}
+(3f-1)f'\frac{a'}{a}+2(d-5)(f-1)\frac{f'}{r }
-(f-1)f\frac{a'^2}{a^2}
+(d-5)(d-6)\frac{(f-1)^2}{r^2}
\bigg)=0,
\end{eqnarray}
 \begin{eqnarray}
\label{eqb} 
&&b''+(\frac{b'}{2}+(d-3)\frac{b}{r})\frac{f'}{f}-\frac{b'^2}{2b}+(d-3)(d-4)\frac{b(f-1)}{r^2 f}
+(d-3)\frac{b'}{r}
\\
\nonumber
&&{~~~~~~~~~~}
-\frac{\alpha}{4}\frac{(d-3)(d-4)b}{r^2 f}
\bigg(
2(f-1)f\frac{b''}{b}+ 2(d-5)(f-1)f\frac{b'}{r b}
\\
\nonumber
&&{~~~~~~~~~~}
+(3f-1)f'\frac{b'}{b}+2(d-5)(f-1)\frac{f'}{r }
-(f-1)f\frac{b'^2}{b^2}
+(d-5)(d-6)\frac{(f-1)^2}{r^2}
\bigg)=0,
\end{eqnarray}
\begin{eqnarray}
\label{eqf} 
&& (d-2)\frac{f'}{rf}
-\frac{a'b'}{2ab}
+\frac{1}{r}(\frac{a'}{a}+\frac{b'}{b})
+(d-1)(d-4)\frac{(f-1)}{r^2 f}
\\
\nonumber
&&{~~ }
+(d-4)\frac{\alpha}{4}
\bigg(
-\frac{(d-5)(f-1)}{r^4 f}
\big(  (d-6)(d+1)(f-1)+2drf' \big)
-\frac{1}{r^3}(\frac{a'}{a}+\frac{b'}{b})
\big( 
6(d-5)(f-1)
\\
\nonumber
&&{~~~~~~~~~~}
+r^2 f\frac{a'b'}{ab}+2r(3f-1)\frac{f'}{f} 
\big)
+\frac{2(f-1)}{r^2}(\frac{a'^2}{a^2}+\frac{b'^2}{b^2})
+\frac{a'b'}{r^2ab}\left(3r f'+(d-5)(3f-1) \right)
\\
\nonumber
&&{~~~~~~~~~~}
+\frac{2a''}{r^2a}\big(rf\frac{b'}{b}-2(f-1) \big)
+\frac{2b''}{r^2b}\big(rf\frac{a'}{a}-2(f-1) \big)
\bigg)
=0,
\\
\label{cons} 
&&-\frac{1}{2}(d-3)(d-4)\frac{f-1}{f}
-(d-3)\frac{r}{2}(\frac{a'}{a}+\frac{b'}{b})
-\frac{r^2 a'b'}{4 ab}
\\
\nonumber
&&{~~~~~~~~~~}
+\frac{\alpha}{4}
\bigg(
(d-5)(d-6)\frac{(f-1)^2}{2r^2f}
+(d-5)\frac{(f-1)}{r}(\frac{a'}{a}+\frac{b'}{b})
+(3f-1)\frac{a'b'}{2ab}
\bigg)
=0,
\end{eqnarray}
However, only three of the above equations are independent and one can show that (\ref{cons}) is
in fact
a differential consequence of (\ref{eqa})-(\ref{eqf}).

The equations under consideration are invariant under the following rescaling:
\be
\label{scaling}
\alpha \rightarrow \lambda^2 \alpha,\ r \rightarrow \lambda r \ .
\ee
A dimensionless relevant parameter can then be defined according to
$\beta = \alpha/\lambda^2 $,
where $\lambda$ is some length scale.
It is natural and convenient to choose $\lambda$ as the horizon radius $r_h$ 
of the black string and thus to define
\be
\label{beta_def}
\beta \equiv \frac{\alpha}{r_h^2}.
\ee
\subsection{Boundary conditions and asymptotic solutions}
We require that our solutions approach asymptotically the background metric (\ref{KK-metric}), which imposes 
  $a(r)=b(r)=f(r)=1$ as $r \to \infty$. 
A black string has also a regular horizon at $r=r_h$, where $f(r_h)=b(r_h)=0$, and $a(r_h)>0$, 
$f'(r_h)>0$, $b'(r_h)>0$.
Restricting to nonextremal configurations,
local solutions in the vicinity of the event horizon $r=r_h$
can be constructed by expanding them in terms of $r-r_h$ as
\begin{eqnarray}
\label{exphor}
a &=& a_0 + a_1(r-r_h)+\Ord{r-r_h}{2},\nonumber \\
b &=& b_1(r-r_h)+ b_2(r-r_h)^2 + \Ord{r-r_h}{3},\\
f &=& f_1(r-r_h)+ f_2(r-r_h)^2 + \Ord{r-r_h}{3}. 
\nonumber 
\end{eqnarray}
All constant in the  near event horizon expansion are expressed in terms of the positive
parameters $a_0,b_1$ (the expressions of $a_1$, $b_2$ and $f_2$ are quite complicated and not enlightening;  
thus we shall not present them here).
 More important is the parameter $f_1$ which is fixed by
\be
\label{f1}
     f_1 = \frac{\beta^2 (3d-14)d_3 d_5 d_4^2 + 12 \beta d_2 d_4 d_5 + 8 d_4 
     - \sqrt{P}}
     {4 \beta r_h (\beta d_3 d_4^2 d_5 + 2 d_4 d_2) },
\ee
with
\be
     P \equiv \beta^4 d_2^2 d_3^2 d_5^2d_4^4 + 8 \beta^3 d_2 d_3^2 d_4^4 d_5 + 32 \beta^2 d_6 d_5 d_4^2 d_2^2
         + 64 \beta d_7 d_4 d_2^2 + 64 d_2^2, \nonumber
\ee
(for shortness, we have used here $d_k \equiv d-k$). 
Note that in the five-dimensional case,  
the obvious condition $P>0$ implies the existence of a minimum horizon size for
a given value of the GB coupling constant\footnote{This is a nonperturbative effect, which cannot be seen considering the GB term 
as a small deformation  of the Einstein gravity.}
 \begin{eqnarray}
\label{ubound}
\frac{\alpha}{r_h^2} \equiv \beta<\frac{1}{2},
\end{eqnarray}
$i.e.$ for $\alpha>0$, one finds $r_h\geq \sqrt{~2\alpha }$.
Since, as discussed in Ref. \cite{Kobayashi:2004hq}, the horizon radius is 
decreasing monotonically with the mass of the solutions,
the relation (\ref{ubound}) shows  the existence 
of a minimal value of the mass for a given 
GB coupling constant $\alpha$ (note that a similar property is found for $d=5$ Schwarzschild-GB
black hole, see Appendix A). 
For $d>5$ there is, as far as we can see, no upper bound on $\beta$.
However, exploring the second derivative of the function $f(r)$ around the horizon reveals the existence of 
another (negative) critical value:
\be
 \beta_c = -\frac{2}{(d-3)(d-4)},
\label{lbound}
\ee
with $f_2$ diverging as $\beta\to \beta_c$.
Although the existence of solutions with $\beta<\beta_c$ cannot be excluded,
we could not find these disconnected branches.
Then we conjecture that, similar to $d=5$ case \cite{Suranyi:2008wc}, 
 for $r_h$ fixed, there exists a lower bound for 
the GB coupling constant, $\alpha>-\frac{2r_h^2}{(d-3)(d-4)}$. 
Seen oppositely, there exists an upper bound on the horizon radius $r_h$ ($i.e.$ for the mass) 
for a fixed  negative value\footnote{Note that string theory requires $\alpha>0$ and solutions with negative $\alpha$
may be unphysical.
Nevertheless,  in the study of UBS we have considered this possibility for the
sake of completness.} of $\alpha$.

For large values of $r$, the functions $a,b,f$ admit the following expansion
in terms of two constants $c_t$, $c_z$
\bea
\label{expas}
&&a(r)=1 + \frac{c_z}{r^{d-4}}+ \frac{c_t c_z}{2r^{2(d-4)}} + \dots,~~~~
b(r)=1 -\frac{c_t}{r^{d-4}}+ \frac{c_t c_z}{2r^{2(d-4)}} + \dots,
\\
&&f(r)=1 +\frac{c_z-c_t}{r^{d-4}} + \frac{d-4}{2(d-3)}\frac{c_t c_z}{2r^{2(d-4)}} + \dots.
\nonumber
\eea
One can see that the GB coupling constant does not appear in the first terms of this expansion. 
However, we have found that asymptotic corrections specific  to the GB interaction
 occur at higher orders in $1/r$.

Obtaining solutions extrapolating from \eqref{exphor} to \eqref{expas} allows 
to determine the values of the parameters $a_0,b_1,c_t$ and $c_z$ as 
functions of $r_h$ and $\alpha$, which are the input 
data in the numerics.

\subsection{A Smarr relation and scaling properties}
The thermodynamic properties of these solutions  are determined
by the asymptotic charges --the mass and tension,
and the quantities on the horizon --the temperature and the entropy.

The Hawking temperature $T_H$ of a black string 
as computed by using 
the  definition (\ref{TH-gen}) in terms of the surface gravity or  
by  demanding the regularity on the Euclidean section is
\be
T_H = \frac{\sqrt{b'(r_h)f'(r_h)}}{4\pi}.
\ee
The entropy $S$ of a uniform black string is computed by using the general relation (\ref{S-Noether}),
\be
\label{entrop1}
S = \frac{V_{d-3}L r_h^{d-3}\sqrt{a(r_h)}}{4G}\left(1+\frac{1}{2}\beta(d-3)(d-4)\right).
\ee
Also, by integrating the identities
\begin{eqnarray}
\label{totder1} 
 &R_t^t+\frac{\alpha}{4}(H_t^t+\frac{1}{2}L_{GB})=
\frac{1}{ r^{d-3}} \sqrt{\frac{f}{ ab }}\frac{d }{dr}
 \left  ( 
   \frac{1}{2}r^{d-3}b'\sqrt{\frac{af}{ b }}(-1+
  \frac{1}{2}\alpha(d-3)
  (
  (d-4)(f-1)+f \frac{ra'}{a}
  )
\right ),~~~~{~~} 
\\
\label{totder2} 
 &R_z^z+\frac{\alpha}{4}(H_z^z+\frac{1}{2}L_{GB})=
\frac{1}{ r^{d-3}} \sqrt{\frac{f}{ ab }}\frac{d }{dr}
 \left  ( 
  \frac{1}{2}r^{d-3}a'\sqrt{\frac{bf}{a }}(-1+
  \frac{1}{2}\alpha(d-3)
  (
  (d-4)(f-1)+f \frac{rb'}{b}
  )
\right ),~~~~{~~} 
\end{eqnarray} 
between the event horizon and infinity, together with the field equations (\ref{eqs}),
one can  prove an unexpectedly simple 
Smarr-type 
formula\footnote{Note that the relation (\ref{totder1})
can be used to derive the expression (\ref{entrop1})
for the entropy of UBS within the Euclidean approach to quantum gravity \cite{Hawking:ig}.
In this case, one has to supplement the action (\ref{action}) with the corresponding boundary terms for
Einstein and Gauss-Bonnet gravity.
The tree level action $I$ is regularized by subtracting the contribution of the background (\ref{KK-metric}).
Then the entropy is computed from $F=M-T_H S$, with the free energy $F=T_H I$, and $M$ given by (\ref{2})
As expected, the entropy computed in this way agrees with (\ref{entrop1}).  
}. 
This formula relates quantities defined at 
infinity to quantities defined at the event horizon:
\begin{eqnarray}
\label{smarrform} 
M(1-n)= T_H S~,
\end{eqnarray} 
which also provides a useful check of the 
accuracy of the numerical solutions\footnote{One can verify that 
this relation is trivially satisfied by the UBSs in Einstein gravity.
However, the $\alpha=0$ string solutions
have another Smarr relation \cite{Harmark:2003dg}
\begin{eqnarray}
\label{smarrform1} 
T_HS =  M\frac{d-3-n}{d-2},
\end{eqnarray}
which holds also for configurations with a dependence on the extra
dimension.
Due to the presence of a new length scale set by $\alpha$,
we could not generalize
this relation  to EGB theory.
}. 
 
The rescaling (\ref{scaling}) 
 affects the thermodynamical quantities in the following way:
\bea
\label{scal-new}
&& T_H \rightarrow \frac{T_H}{\lambda},\ S\rightarrow \lambda^{d-3}S\nonumber\\
&& M \rightarrow \lambda^{d-4}M,\ \mathcal T \rightarrow \lambda^{d-4}\mathcal T,
\eea
while $\alpha\to \lambda^2 \alpha $, $r_h\to \lambda r_h$. 

We close this part by remarking that
all UBSs in this work
have an alternative interpretation as bubble solutions.
The bubbles are found by using
the analytic continuation $z \to i u $, $t \to i\chi$ 
in the general line element (\ref{ansatz}):
\begin{eqnarray}
\label{metric-b} 
ds^2=-a(r)du^2+b(r)d\chi^2+ \frac{dr^2}{f(r)}
+r^2d\Omega^2_{d-3},
\end{eqnarray}
(where $\chi$ has a periodicity $\beta_\chi=1/T_H$).
The properties of the bubbles
can be discussed by using similar methods.
For example, their mass and tension are fixed by the tension and mass of the corresponding black strings.

\subsection{Explicit solutions}
To our knowledge, for  $\alpha\neq 0$, the equations (\ref{eqa})-(\ref{eqf}) have no general closed form solutions.
However, one can analyse the 
properties of the EGB uniform strings by using a combination of analytical and numerical
methods, which is enough for most purposes.

\subsubsection{Einstein gravity UBSs}
The system of equations (\ref{eqa})-(\ref{cons}) admit several exact solutions in particular limits.  
We start with $\alpha = 0$ solutions, in which case the family of 
$(d-1)$-dimensional black hole solutions of Schwarzschild-Tangherlini 
 can be uplifted in $d$-dimensions:
\be
\label{Einstein-UBS}
      f(r) = b(r) = 1 - \left(\frac{r_h}{r}\right)^{d-4}\equiv f_0(r)\ \ , \ \ a(r) = 1,
\ee
$r=r_h$ corresponding to the event horizon radius. 
The mass, tension, Hawking temperature and entropy of the Einstein  gravity UBSs
can be written in terms of $r_h$ as
\be
\label{rel-E}
  M^{(E)}=\frac{LV_{d-3}(d-3)r_h^{d-4}}{16\pi G},~~ 
  {\cal T}^{(E)}=\frac{V_{d-3}r_h^{d-4}}{16\pi G},~~
  T_H^{(E)}=\frac{(d-4)} {4\pi r_h},~~
   S^{(E)}=\frac{LV_{d-3} r_h^{d-3}}{4 G}. 
\ee
These black strings exist for all values of $r_h>0$
and have a constant relative tension $n=1/(d-3)$.
Other properties of these solutions are similar to those of the $(d-1)$-dimensional 
Schwarzschild black holes.
For example, their entropy decreases with the temperature.
Thus the static Einstein gravity UBSs are thermally unstable. 

\subsubsection{The small $\alpha$-solutions}
An exact solution of the system (\ref{eqa})-(\ref{eqf})
can be found in the limit of small $\alpha$ by treating
the EGB configurations as a perturbation around the Einstein gravity.
Here we have found convenient to take an ansatz with:
\bea
\label{eq-fo}
&&a(r)=1+\alpha a_1(r)+ \alpha^2 a_2(r) +\dots,~~b(r)=f_0(r)\left(1+\alpha b_1(r)+ \alpha^2 b_2(r)+ \dots \right) ,
\\
\nonumber
&&f(r)=f_0(r)\left(1+\alpha f_1(r)+ \alpha^2 f_2(r)  +\dots\right).
\eea
In the first order in $\alpha$ one arrives at the system of linear ordinary 
differential equations
\begin{eqnarray}
\nonumber
&&f_0 a_1''+\frac{2(d-3)-(d-2)(\frac{r_h}{r})^{d-4}}{2r}a_1'
+\frac{d-3}{r}f_0 f_1'
+\frac{(d-3)(d-4)}{r^2 }f_1
\\
\nonumber
&&{~~~~~~~~~~~~}+\frac{1}{4r_h^4 }~\left(\frac{r_h}{r}\right)^{2d-4}(d-2)(d-3)(d-4)(d-5)=0,
\\
\label{pert-alpha-eqs} 
&&f_0 b_1''
+\frac{(2(d-3)+(d-6)(\frac{r_h}{r})^{d-4})}{2r}b_1'
+\frac{(2(d-3)-(d-2)(\frac{r_h}{r})^{d-4})}{2r }f_1' 
\\
\nonumber
&&{~~~~~~~~}
+\frac{(d-3)(d-4)}{r^2} f_1
-(d-2)(d-3)^2(d-4)\frac{1}{4r^4}(\frac{r_h}{r})^{2(d-4)}=0,
\\
\nonumber
&& \left(d-4+(d-2)f_0 \right)a_1'
+2(d-3)f_0 b_1'
+\frac{2(d-3)(d-4)}{r}f_1
\\
\nonumber
&&{~~~~~~~~}+(d-2)(d-3)(d-4)(d-5)\frac{1}{2r^3}(\frac{r_h}{r})^{2(d-4)}=0.
\end{eqnarray}
Similar equations are found for higher order functions, 
in which case they have a much more complicated form (and therefore are not shown here).  
When solving the equations for some order $k$ in perturbation theory, 
there are four integration constants.
These constants are choosen such that the corrected UBS metric still has an horizon
at $r=r_h$ and approaches the background (\ref{KK-metric}) asymptotically.

Unfortunately, we could not find a general solution 
  valid for any $d$ and
the equations has to be solved 
separately for any spacetime dimension.
For example, in the first order in $\alpha$, the solution reads
\begin{eqnarray}
\label{d=5}
a_1(r)=
\frac{2}{3r_h r}
+\frac{1}{3r^2}
+\frac{2r_h}{9r^3},
~b_1(r)=-\frac{23}{36r_hr}-\frac{11}{36r^2}-\frac{ r_h}{6r^3} ,
~f_1(r)=\frac{1}{36r_hr}+\frac{7}{36r^2}+\frac{5 r_h}{18r^3},~~{~~~~}
\end{eqnarray}
for $d=5$, and
\begin{eqnarray}
\label{d=6}
a_1(r)=\frac{9}{8r^2}+\frac{9r_h^2}{16r^4} ,~~b_1(r)=-\frac{23}{16 r^2}-\frac{7r_h^2}{8r^4} ,
~~f_1(r)=-\frac{5}{16 r^2}+\frac{r_h^2}{16r^4} ,
\end{eqnarray}
in six spacetime dimensions.
The expression of these functions becomes more complicated for 
$d>6$. For example, the solution for $d=7$ is
\begin{eqnarray}
\nonumber
&&a_1(r)= \frac{4}{25}
\bigg(
\frac{5\sqrt{3}\pi}{r_h^2}
+\frac{15}{r^2}
+\frac{6r_h^3}{r^5}
-\frac{10\sqrt{3}}{r_h^2}\arctan(\frac{2r+r_h}{\sqrt{3}r_h})
-\frac{15}{r_h^2}\log (1+\frac{r_h}{r}+(\frac{r_h}{r})^2),
\\
\label{d=7}
&&b_1(r)= \frac{1}{5f_0(r)}
\bigg(
-\frac{ \sqrt{3}\pi}{ r_h^2}
-\frac{3 }{  r^2}
+\frac{r_h(-138 +20\sqrt{3}\pi+45\log 3)}{10r^3}
+\frac{63r_h^3}{10r^5}
\\
\nonumber
&&{~~~~~~}
+\frac{21r_h^6}{2r^8}
+
\frac{\sqrt{3}}{r_h^2}
 (2-\frac{5r_h^3}{r^3})
 \big(
 \arctan(\frac{2r+r_h}{\sqrt{3}r_h})
+\frac{\sqrt{3}}{2}\log (1+\frac{r_h}{r}+(\frac{r_h}{r})^2)
 \big)
\bigg),
\\
\nonumber
&&f_1(r)= \frac{1}{5f_0(r)}\bigg(
\frac{(10\sqrt{3}\pi-78+45\log 3)}{10r^3}
+\frac{9r_h^3}{2r^5}
+\frac{33r_h^6}{10r^8}
\\
\nonumber
&&{~~~~~~~~~~~~~~~~}
-\frac{3\sqrt{3}r_h}{ r^3}
  \big(
\arctan(\frac{2r+r_h}{\sqrt{3}r_h})
+\frac{\sqrt{3}}{2}\log (1+\frac{r_h}{r}+(\frac{r_h}{r})^2)
 \big)
\bigg ),
\end{eqnarray}
while for $d=8$ one finds
\begin{eqnarray}
\label{d=8}
&&a_1(r)= 
\frac{25}{6r^2}
+\frac{25r_h^4}{18r^6}
-\frac{25}{6r_h^2}\log(1+(\frac{r_h}{r})^2),
\\
\nonumber
&&b_1(r)=
\frac{1}{f_0(r)}
\bigg (
-\frac{5}{6r^2}
+\frac{ r_h^2(15\log 2-47)}{9r^4}
+\frac{20r_h^4}{9r^6}
+\frac{23r_h^8}{6r^{10}}
+(1-3(\frac{r_h}{r})^4)\frac{5}{6r_h^2}\log(1+(\frac{r_h}{r})^2)
\bigg ),
\\
\nonumber
&&
f_1(r)=
\frac{1}{f_0(r)}
\bigg (
\frac{r_h^2(-32+15\log 2)}{9 r^4}
+\frac{5r_h^4}{3r^6}
+\frac{17r_h^8}{9r^{10}}
-\frac{5r_h^2}{3r^4}\log(1+(\frac{r_h}{r})^2)
\bigg ).
\end{eqnarray}
Similar expressions have been found for $d=9,10$, 
without being possible to identify
a general pattern\footnote{The presence of $\log r$ and $\arctan (h_1 r+h_2)$
terms ~(with $h_1,h_2$ dimension dependent parameters) is a generic feature of the $d>6$ solutions.}.

This leads to following expression of the relevant quantities exhibiting
first order corrections in $\beta=\alpha/r_h^2$:
\begin{eqnarray}
\label{122}
M=M^{(E)}(1+c_1\beta),~~{\cal T}={\cal T}^{(E)}(1-c_2\beta),~~T_H=T_H^{(E)}(1-c_3\beta),~~
S=S^{(E)}(1+c_4\beta),~{~~}
\end{eqnarray}
where $c_i$ are dimension-dependent positive numbers, $e.g.$:
\begin{eqnarray}
\nonumber
&&c_1=c_3=\frac{11}{36},~~
c_2=\frac{25}{36},~~
c_4=\frac{29}{18},~~~{\rm for}~~d=5,
\\
\label{121}
&&c_1=\frac{17}{16},~~
c_2=\frac{31}{16},~~
c_3=\frac{41}{32},~~
c_4=\frac{123}{32},~~~{\rm for}~~d=6,
\\
\nonumber
&&c_1=\frac{ 138+5\sqrt{3}\pi -45\log 3}{50},~~
c_2=\frac{  5\sqrt{3}\pi -9(18+5\log 3)}{50},~~
c_3=\frac{438+ 5\sqrt{3}\pi -45\log 3}{150},~~
\\
\nonumber
&&{~~~~~~~~~~~~~~~~~~~~~~~~~~~~~~~~~~~~~~~~}
c_4=\frac{2(288+5\sqrt{3}\pi -45\log 3}{75},~~~~~{\rm for}~~d=7,
\\
&&
\nonumber
c_1=\frac{47-15\log 2}{9},~~
c_2=\frac{43+15\log 2}{9},~~
c_3=\frac{182-15\log 2}{36},~~
c_4=\frac{5(92-15\log 2)}{36},~~{\rm for}~~d=8,
\end{eqnarray}
$M^{(E)}$, ${\cal T}^{(E)}$, $T_H^{(E)}$ and $S^{(E)}$
being the corresponding quantities in the Einstein gravity as given by (\ref{rel-E}).

One can see that, in the first order perturbation theory, 
the mass and entropy increase with $\beta$ 
while Hawking temperature and tension decrease.

However, the situation is different when considering
higher order corrections.
For example, in the third order one finds the following expressions for the relevant quantities\footnote{The form 
of the metric functions are rather similar to (\ref{d=5}), (\ref{d=6}).}:
\begin{eqnarray}
\label{123}
&&M=M^{(E)}(1+\frac{11}{36}\beta+\frac{127}{600}\beta^2+\frac{6922513}{76204800}\beta^3),~
{\cal T}={\cal T}^{(E)}(1-\frac{25}{36}\beta+\frac{67}{600}\beta^2+\frac{7796449}{76204800}\beta^3),~~{~~}
\\
\nonumber
&&T_{H}=T_H^{(E)}(1-\frac{11}{36}\beta+\frac{2737}{32400}\beta^2-\frac{2650383}{228614400}\beta^3),~
S=S^{(E)}(1+\frac{29}{18}\beta+\frac{23311}{32400}\beta^2+\frac{10632569}{38102400}\beta^3),~{~~}
\end{eqnarray}
for $d=5$, while the corresponding $d=6$ expressions are
\begin{eqnarray}
\label{124}
&&M=M^{(E)}(1+\frac{17}{16}\beta-\frac{125}{3072}\beta^2-\frac{223547}{307200}\beta^3),~
{\cal T}={\cal T}^{(E)}(1-\frac{31}{16}\beta+\frac{8995}{3072}\beta^2-\frac{837467}{307200}\beta^3),~~~~{~~~~}
\\
\nonumber
&&T_{H}=T_H^{(E)}(1-\frac{41}{32}\beta+\frac{6487}{3072}\beta^2-\frac{9824347}{2457600}\beta^3),~
S=S^{(E)}(1+\frac{123}{32}\beta+\frac{1319}{1024}\beta^2-\frac{1799983}{819200}\beta^3).
\end{eqnarray} 
It is easy to verify that the above expressions verify the first law (\ref{firstlaw}) and
the Smarr relation (\ref{smarrform}) up to order $\Ord{\beta}{4}$.

One should also note that, while for $d=5$ the mass and entropy always increase with 
$\beta$, the situation is different for $d=6$.
Also, the above results make clear that the relative tension $n$
cannot be constant  for EGB black strings, since in the third order one finds
\begin{eqnarray}
n= \frac{1-\frac{25}{36}\beta+\frac{67}{600}\beta^2+\frac{7796449}{76204800}\beta^3}
{2(   1+\frac{11}{36}\beta+\frac{127}{600}\beta^2+\frac{6922513}{76204800}\beta^3)} 
~~~~~{\rm for}~~~~~d=5,
\end{eqnarray} 
and
\begin{eqnarray}
n= \frac{1-\frac{31}{16}\beta+\frac{8995}{3072}\beta^2-\frac{837467}{307200}\beta^3}
{3(1+\frac{17}{16}\beta-\frac{125}{3072}\beta^2-\frac{223547}{307200}\beta^3)} ~~~~~{\rm for}~~~~~d=6.
\end{eqnarray} 
The comparison of the relations (\ref{d=5})-(\ref{121})
with those in Appendix A shows that, for small $\alpha$, one can indeed
treat an EGB black string as an uplifted EGB black hole. 
For example, in the first order perturbation theory, 
one finds the same qualitative dependence on $\alpha$, with different coefficients, however.

\subsubsection{The large $\alpha$-limit. A fixed point analysis}

In the limit where $\alpha\rightarrow\infty$ (which is equivalent to taking only the GB part
in the action), 
particular families of solutions can be constructed with the form 
$a=r^q$, $b=r^p$, $f=f_0$ where $f_0,p,q$ are constants.
In fact, the underlying system of equations can be expressed in terms of $A\equiv ra'/a$, $B \equiv rb'/b$ and 
$f(r)$ and can be considered as a dynamical system in $A,B,f$ and in the variable $\tau= \log(r)$. 
Such systems admit usually  one or several fixed-points.
For example, in the pure Einstein-case, the equations 
can be set is the form
\bea
&&rf' = 2(d-4)\left(1-f\right) - f(r)\left(A+B\right),~~rB'f=  - (d-4)B,~
\\
&&\left(AB + 2(d-3)(A+B)\right. -\left. 2(d-3)(d-4)\right)f = 2(d-3)(d-4),
\nonumber
\eea
 and admit only one fixed point corresponding to 
the flat space-time (\ref{KK-metric}). 
The corresponding equations for the GB theory are more complicated and we shall not present them here.

For $d=5$, these equations have $p=q=0, f_0=1$  as the only  fix-point. 
It  corresponds to Kaluza-Klein space-time (\ref{KK-metric});
obviously, this solution holds for all $d$.
However, for $d>5$, several solutions are available.
More specifically, for $d=6$, we find two more solutions, $q=-1,~p=2, ~f=0$ 
and $a(r)=0,  ~p=2,~ f=0$.
The second is not a fix-point but nevertheless a (singular) solution. 
Note that   $b(r) \sim r^2$ in both cases.

In the case $d=8$, we find more possibilities:
\bea
\nonumber
q&=&0,\ p=-2,\ f=1~~~~
q=q,\ p=-4,\ f=-\frac{1}{3},~~~~
q=0,\ p=-1,\ f=1,
\\
\label{fpd8}
q&=&-\frac{405}{68},\ p=-\frac{9}{2},\ f=-\frac{16}{29},~~~
q=-\frac{5}{2},\ p=-\frac{5}{2},\ f=\frac{8}{3},
\\
\nonumber
q &\approx& 0.415,\ p=-5+\sqrt{13}\approx-1.39,\ f\approx 0.86,~~
q \approx -3.82,\ p=-5-\sqrt{13}\approx-8.6,\ f\approx -0.28.
\eea
A similar list exist for $d=7$ and we expect the same situation for $d>8$.
One crucial difference between the $d=6$ and $d=8$ cases is that
all the solutions (\ref{fpd8}) have $p < 0$ implying that
$b(r)$ is divergent for $r \rightarrow 0$. 
The Ricci scalar is also singular at the origin.

One may expect that
if global solutions ($i.e.$ for $r\in [0,\infty]$)  of the full EGB equations exist  
in the limit $\alpha\rightarrow\infty$,
these solutions would approach one of the fixed points above for $r \to 0$, and the
Kaluza-Klein space-time (\ref{KK-metric}) for $r\to \infty$.

\subsection{Numerical solutions}
\label{sec:ubs}
The perturbative solutions above are useful since may provide some hints 
about the properties of solutions. However, this approach have some clear limits
($e.g.$ it fails to predict the $\beta$-bounds (\ref{ubound}), (\ref{lbound})).

Therefore we rely on numerical methods to construct the  solutions within a nonperturbative approach. 
In this work, we have integrated the system of coupled non linear ordinary differential equations 
with appropriate boundary conditions by using a standard solver \cite{colsys}.
This solver involves a Newton-Raphson method for 
boundary-value ordinary
differential equations, equipped with an adaptive mesh selection procedure.
Typical mesh sizes include $10^2-10^3$ points.
The UBS solutions have a typical relative accuracy of $10^{-6}$. 

\subsubsection{Pattern of solutions}

On general grounds, one  expects a black string to share some of the properties of the corresponding
EGB black holes in $d-1$ dimensions.
Therefore, it follows that
the cases $d=5$ and $d=6$ are special, since the GB term does not contribute in four dimensions, while the $d=5$
EGB black holes  have  distinct properties (see Appendix A). 
Moreover, one can see that some terms vanish in the equations (\ref{eqa})-(\ref{cons}) precisely
for these values of $d$. 

\begin{figure}[ht]
\hbox to\linewidth{\hss%
	\resizebox{8cm}{6cm}{\includegraphics{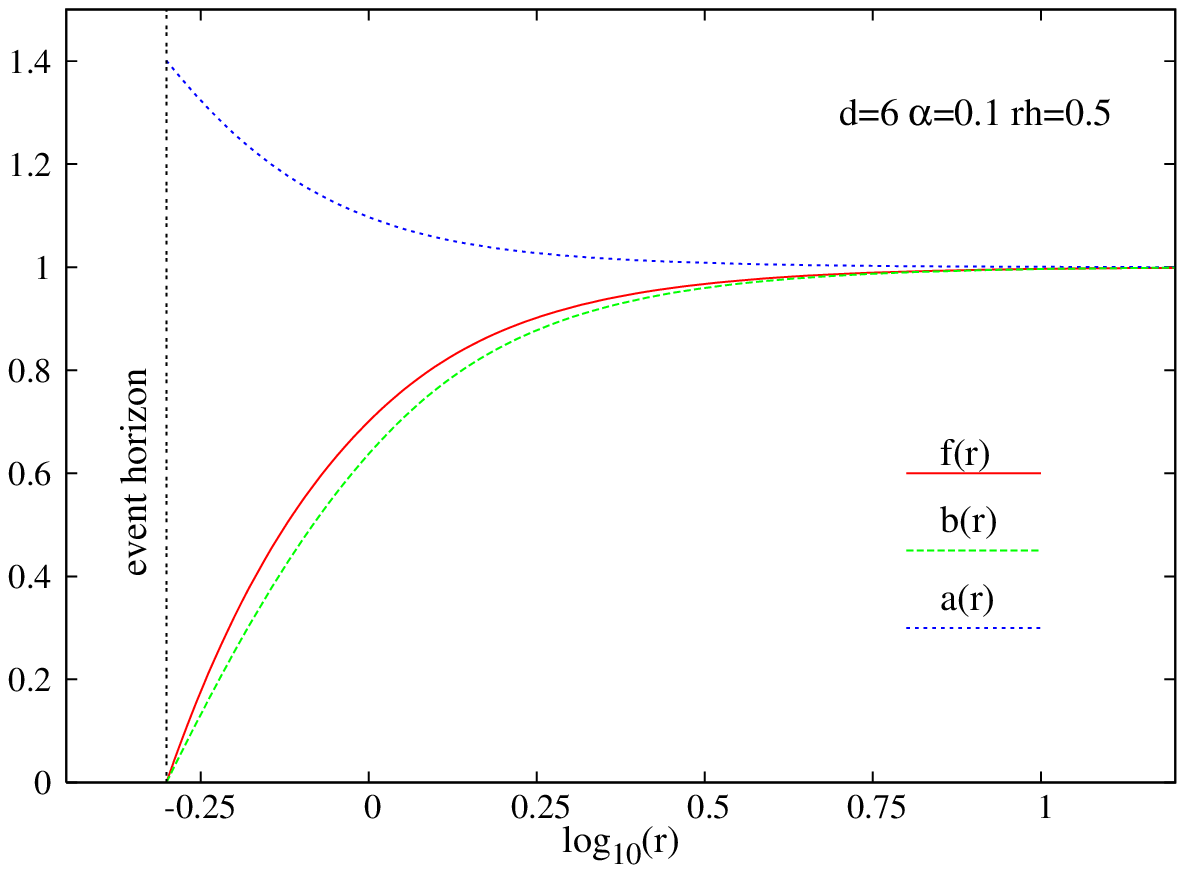}}
\hspace{5mm}%
        \resizebox{8cm}{6cm}{\includegraphics{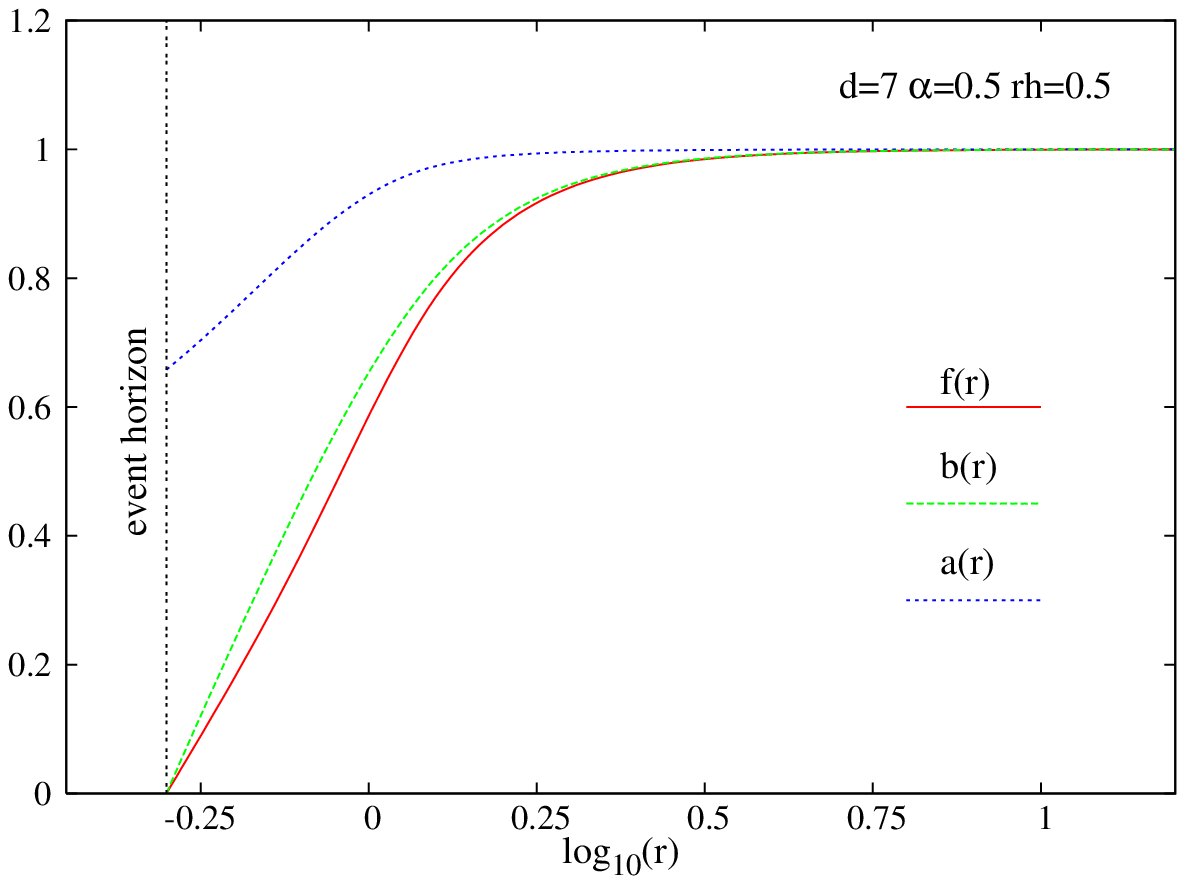}}	
\hss}

\caption{
{\small
 The profiles of the metric functions $-g_{tt}=b(r)$, $g_{zz}=a(r)$ and 
$g^{rr}=f(r) $ are shown for typical $d=6,7$ black string solutions in EGB theory. }
}
\label{profiles}
\end{figure} 

 The case $d=5$ has been studied in \cite{Suranyi:2008wc}; our results in this case 
are in agreement with those of \cite{Suranyi:2008wc} providing a good crosscheck
of the numerical methods.
The distinguished feature of the $d=5$ is the existence of a finite range for the
dimensionless parameter $\beta$, with $-1<\beta<1/2$, with specific 
configurations approached at the limits of the $\beta$-interval.
However, the shape of the metric functions and the qualitative properties are similar to that found for $d>5$.
Therefore, for the rest of this part we shall focus on $d>5$ solutions.

Also, although most of the numerical data presented here corresponds 
to $d=6,8$,  
we have constructed solutions  also for $d=7,9,10$ (however, we did not study all these cases 
in a systematic way).
Therefore we conjecture that UBSs in EGB theory exist for any $d\geq 5$ 
and we hope that our results 
catch the main features of the solutions available for an arbitrary $d$.

For all configurations we have studied,  $a(r)$, $b(r)$ 
and $f(r)$ are smooth functions interpolating  between the corresponding values 
at $r=r_h$ and the asymptotic values at infinity.
A nonzero $\alpha$ leads to 
a deformation of the metric functions of a Einstein gravity black string
at all scales
and is  particularly apparent on the function $g_{zz}=a(r)$ (note that this function 
may possess local extrema for some $r>r_h$).
As expected, for small values of $\beta$, we have noticed a very good agreement
of our results  
with the perturbative solutions (\ref{d=5})-(\ref{d=8}).
The profiles of the metric functions of  typical EGB  black string solutions
are presented on Figure \ref{profiles}.
\begin{figure}[ht]
\hbox to\linewidth{\hss%
	\resizebox{8cm}{6cm}{\includegraphics{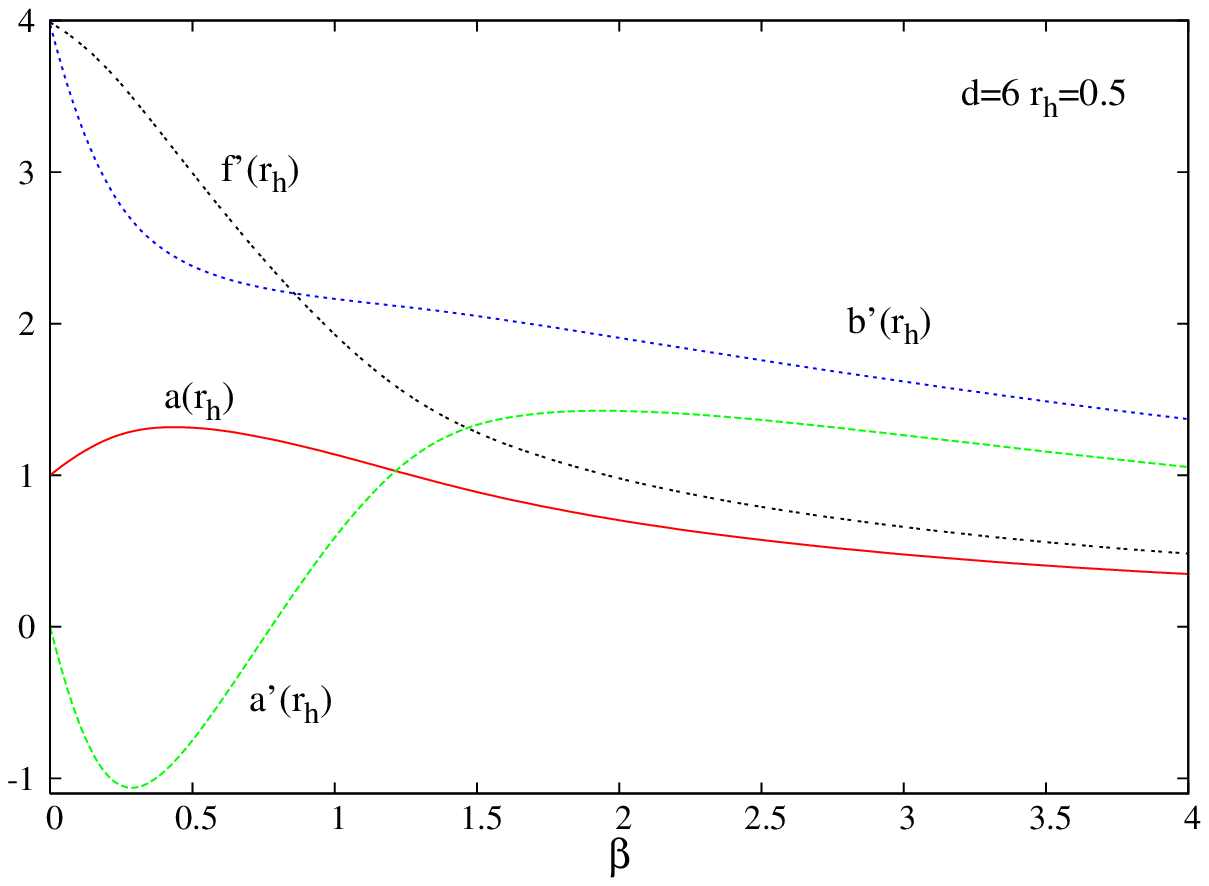}}
\hspace{5mm}%
        \resizebox{8cm}{6cm}{\includegraphics{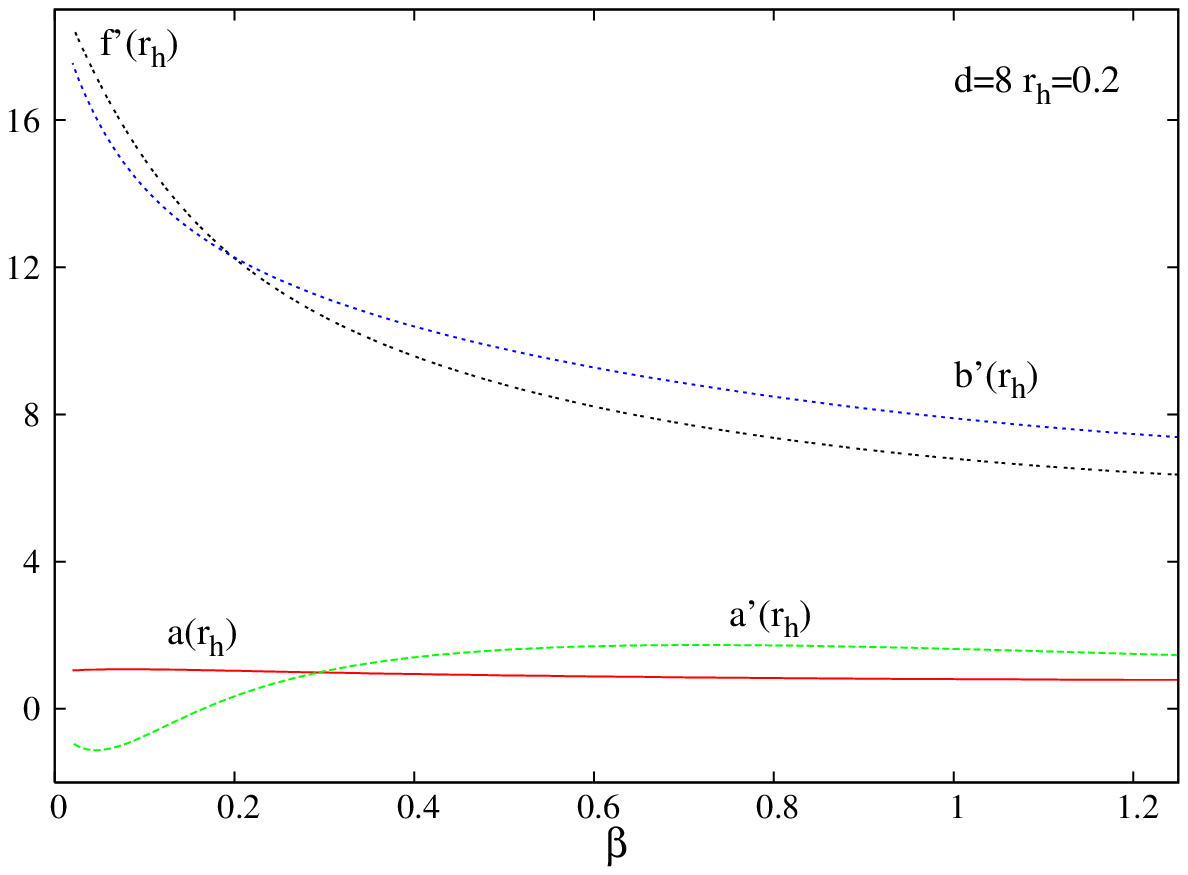}}	
\hss}

\caption{
{\small
The values of the metric function $g_{zz}$ an the horizon, $a(r_h)$, and 
the derivatives of the metric functions at the horizon 
$a'(r_h),b'(r_h),f'(r_h)$ are shown as functions of
the dimensionless parameter  $\beta=\alpha/r_h^2$ for  $d=6,8$ dimensions.}
}
\label{var-alpha-brut}
\end{figure}
 
In practice, for a given value of $d$, we have constructed the full branch of UBS solutions 
both by setting $\alpha=1$ and varying the horizon radius $r_h$ and by taking  $r_h=1$ and varying the GB coupling constant $\alpha$. 
(Note that because of the scaling properties (\ref{scaling}),
these two approaches are equivalent.)
Also, for UBS solutions, the period $L$ of the 
$z$-direction
is an arbitrary positive constant and plays no role in 
 our results.

\begin{figure}[ht]
\hbox to\linewidth{\hss%
	\resizebox{8cm}{6cm}{\includegraphics{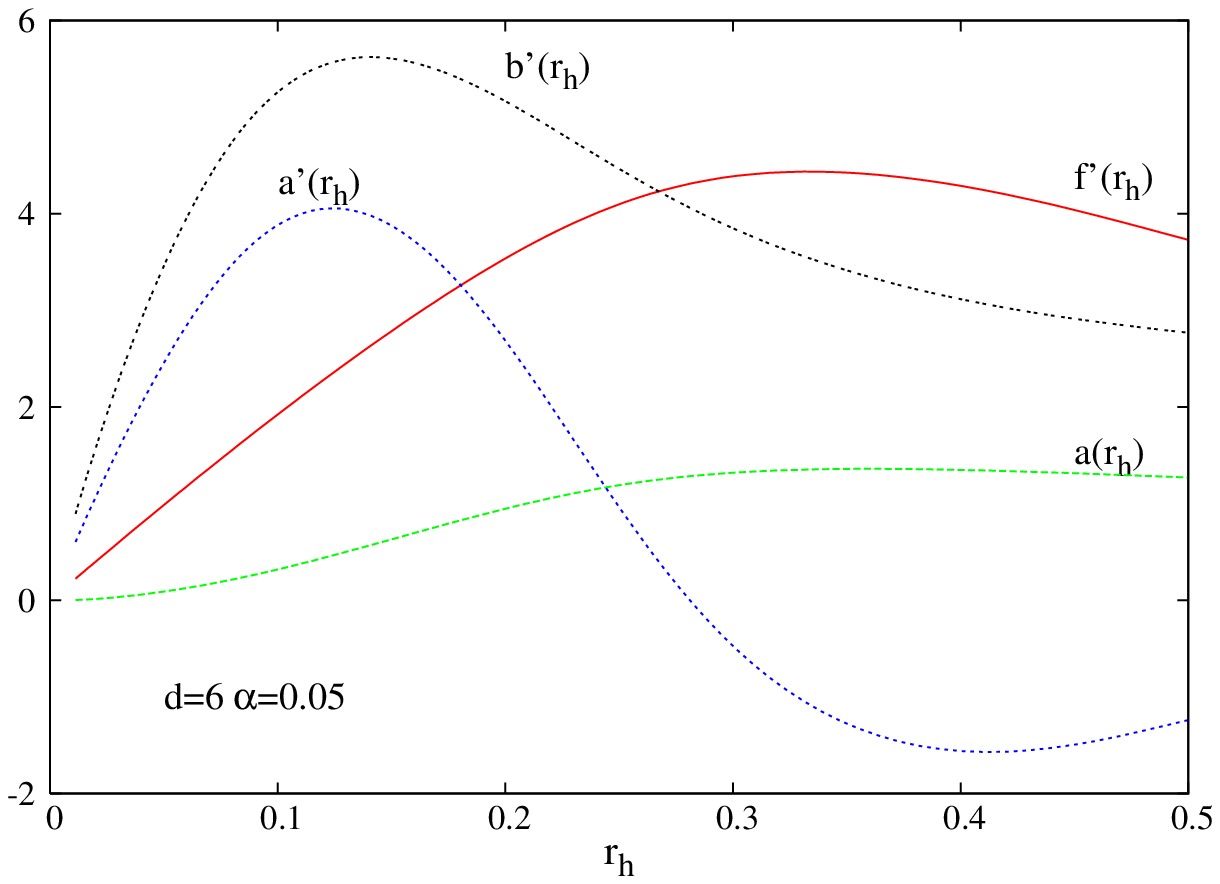}}
\hspace{5mm}%
        \resizebox{8cm}{6cm}{\includegraphics{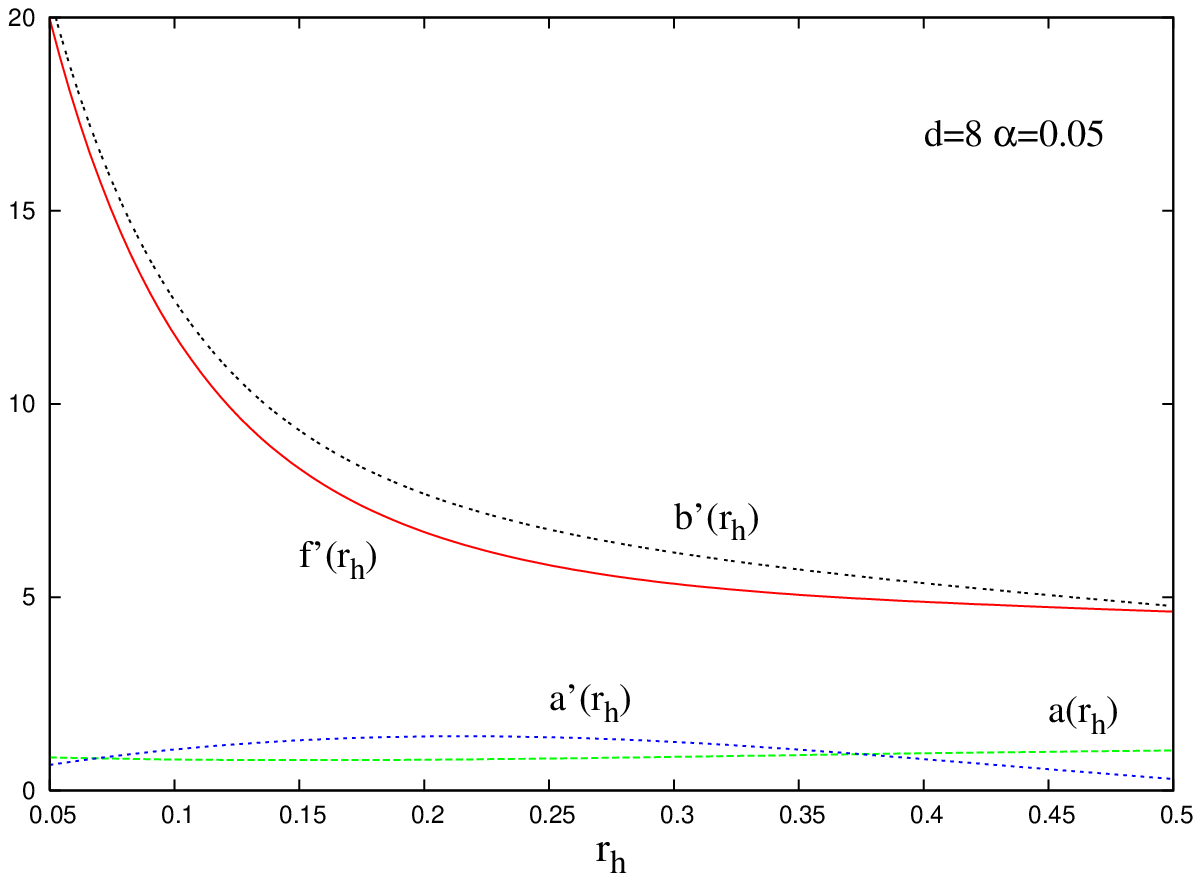}}	
\hss}

\caption{
{\small
The values of the metric function $g_{zz}$ an the horizon, $a(r_h)$, and 
the derivatives of the metric functions at the horizon 
$a'(r_h),b'(r_h),f'(r_h)$, are shown as functions of
the event horizon radius  $r_h$ for  $d=6,8$ dimensions. }
}
\label{var-rh-brut}
\end{figure} 

As can be expected, when the value of the GB coupling is close to the critical (negative) value \eqref{lbound}, 
the solver fails to provide reliable solutions. 
Also, for $d > 5$ we found no signal of a maximal value of $\alpha$ and of $r_h$ 
which may limit the domain of existence of black strings. 
For $r_h$ fixed and in the limit $\alpha \to 0$ (with both signs), 
the numerical solution smoothly approach the Einstein gravity UBS (\ref{Einstein-UBS}).

\begin{figure}[ht]
\hbox to\linewidth{\hss%
	\resizebox{8cm}{6cm}{\includegraphics{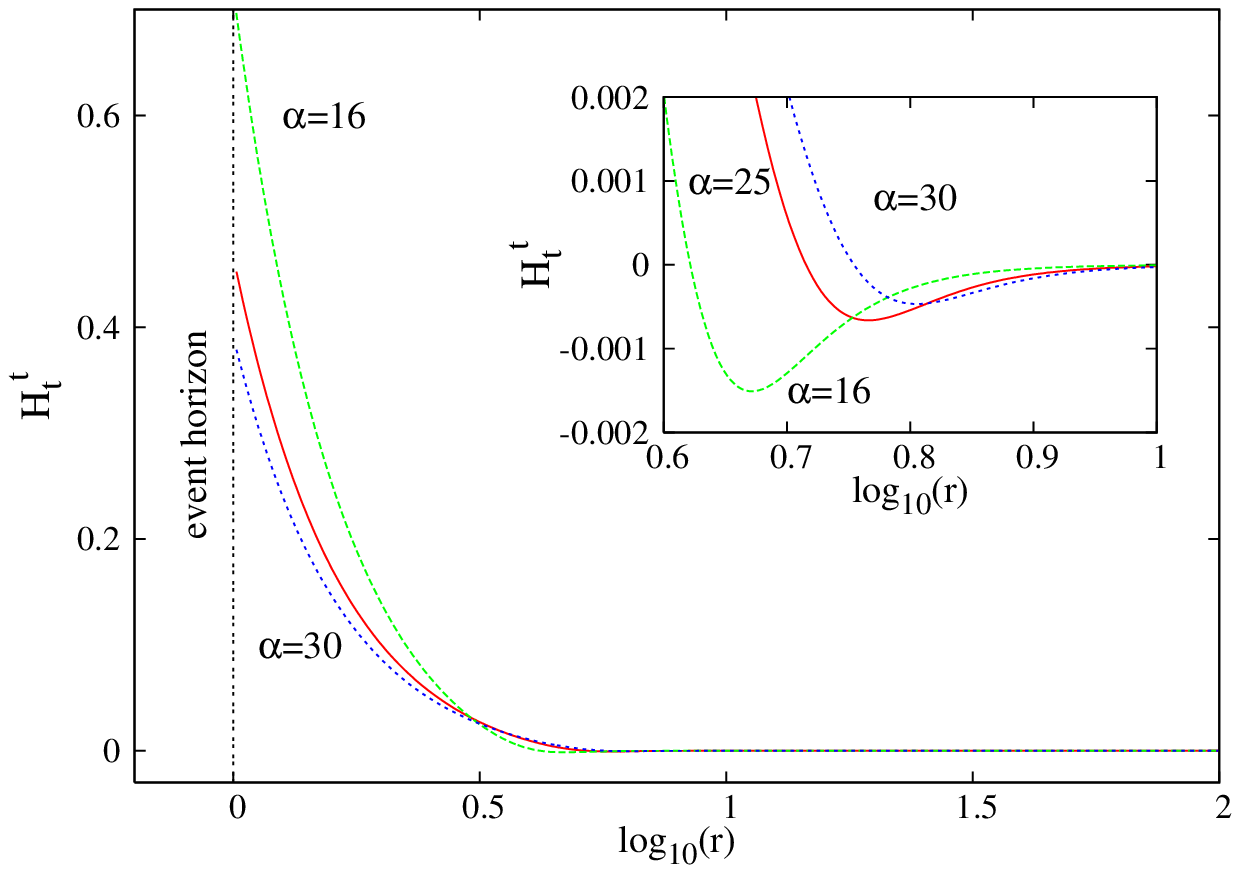}}
\hspace{5mm}%
        \resizebox{8cm}{6cm}{\includegraphics{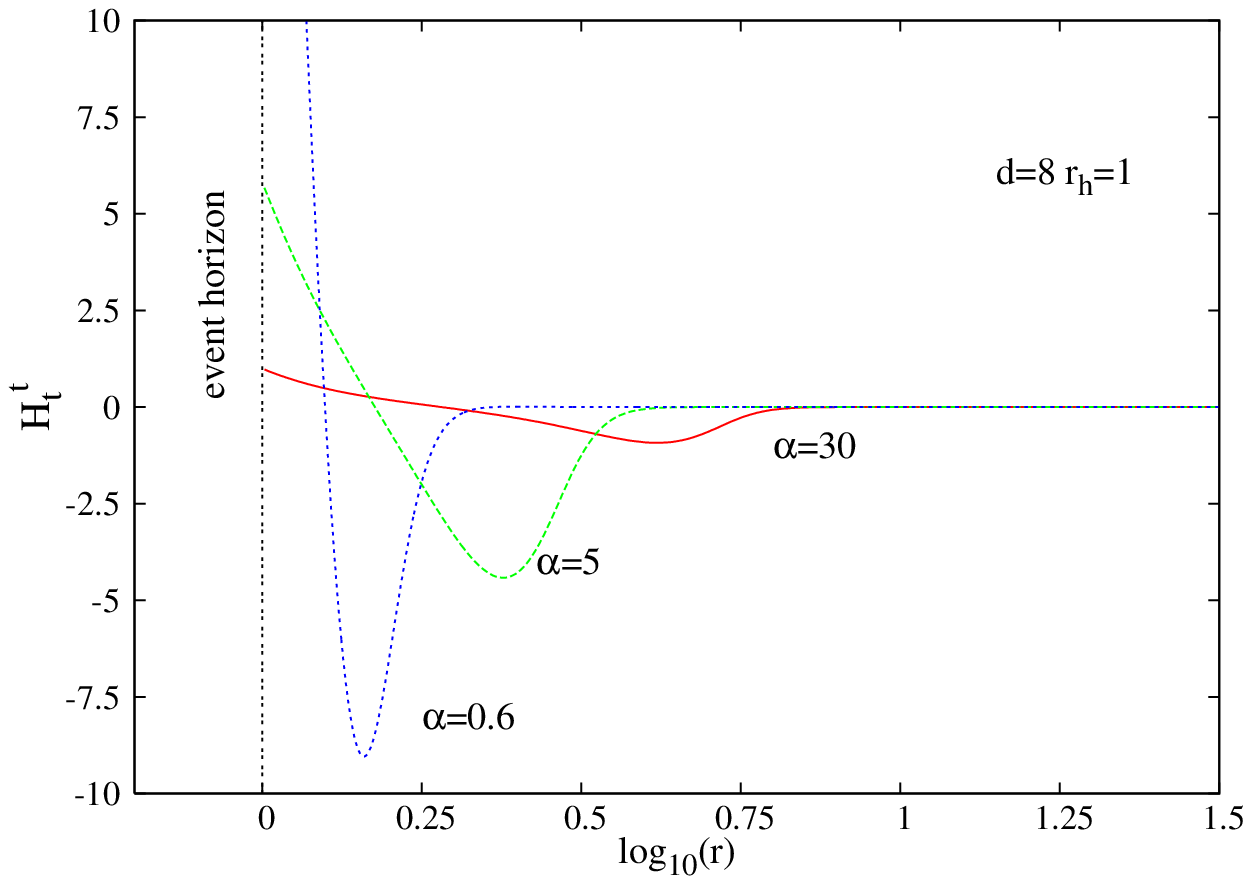}}	
\hss}

\caption{
{\small
The $H^t_t$ component of the Lanczos tensor corresponding to a local  `effective energy density'
is plotted for $d=6,8$ UBS solutions and several values of $\alpha$.}
}
\label{Htt}
\end{figure}
 
 From the point of view of numerics, the solutions can be characterized by 
 the values of the event horizon  parameters $a(r_h), a'(r_h), f'(r_h)$ and $b'(r_h)$
 (although only $a(r_h)$ and $b'(r_h)$ are independent quantities). 
 On Figure \ref{var-alpha-brut}, these parameters 
 are represented in function 
of the dimensionless parameter $\beta$ for $d=6$ (where $r_h=0.5$) and $d=8$ ($r_h=0.2$). 
It can be noticed that the dependence of these parameters on $\alpha$ is not always monotonic.
Also, one can see that for a given $(r_h,\alpha)$,
a UBS solution with the proper asymptotics
 could be found for a single set of the horizon parameters $a(r_h),b'(r_h)$ only.
 
Perhaps more relevant is the study of the branch of solutions obtained with varying $r_h$ and fixed $\alpha$. 
The study of the limit $r_h \to 0$ needs a special attention since it cannot be approached by working with a finite $\beta$.
Our analysis of the black string in this limit reveals a completely different behaviour in the cases $d=6$ and $d=8$.
As shown on Fig. \ref{var-rh-brut} (left), for $d=6$,  the values $a(r_h), a'(r_h), f'(r_h)$ and $b'(r_h)$ all tend to zero for $r_h \to 0$.
This suggest that the black string approach a configuration with a double zero of $f(r)$, $a(r)$ and $b(r)$ at $r=0$.
The limit in the case $d=8$ is easier to interpret: 
the metric approach the Kaluza-Klein background (\ref{KK-metric}) on $]0,\infty]$.
As shown on Fig.  \ref{var-rh-brut} (right) the quantities $f'(r_h), b'(r_h)$ both tend to infinity for $r_h \to 0$, 
indicating that the functions $f(r),b(r)$ becomes discontinuous at the origin and that the flat metric (\ref{KK-metric})
is approached by the black string.
Also, from the data in Figures  \ref{var-alpha-brut}, \ref{var-rh-brut}
one can notice the existence of solutions with $g_{zz}(r_h)>1$ 
and also with $g_{zz}(r_h)<1$, for both $d=6$ and $d=8$.

An explanation of the difference between the $d=6$ and the $d=8$ cases could be 
attempted by inspecting the corresponding fixed points
and remembering that the study of the $r_h\rightarrow0$ limit 
is equivalent to the study of the $\alpha\rightarrow\infty$ limit,
 because of the scaling relations \eqref{scaling}. 

In the case $d=6$, a fixed point exists such that $a=0,f=0,b=r^2$
and the black string solution seems to be attracted by this fixed point in the region of the horizon
for sufficiently small $r_h$. 
The limiting solution for $r_h=0$ seem to extrapolate between the singular fix point
for $r\to 0$ and the background metric (\ref{KK-metric}) for $r\to \infty$.
 This scenario is strongly suggested by Figure  \ref{var-rh-brut} (left).

\begin{figure}[ht]
\hbox to\linewidth{\hss%
	\resizebox{8cm}{6cm}{\includegraphics{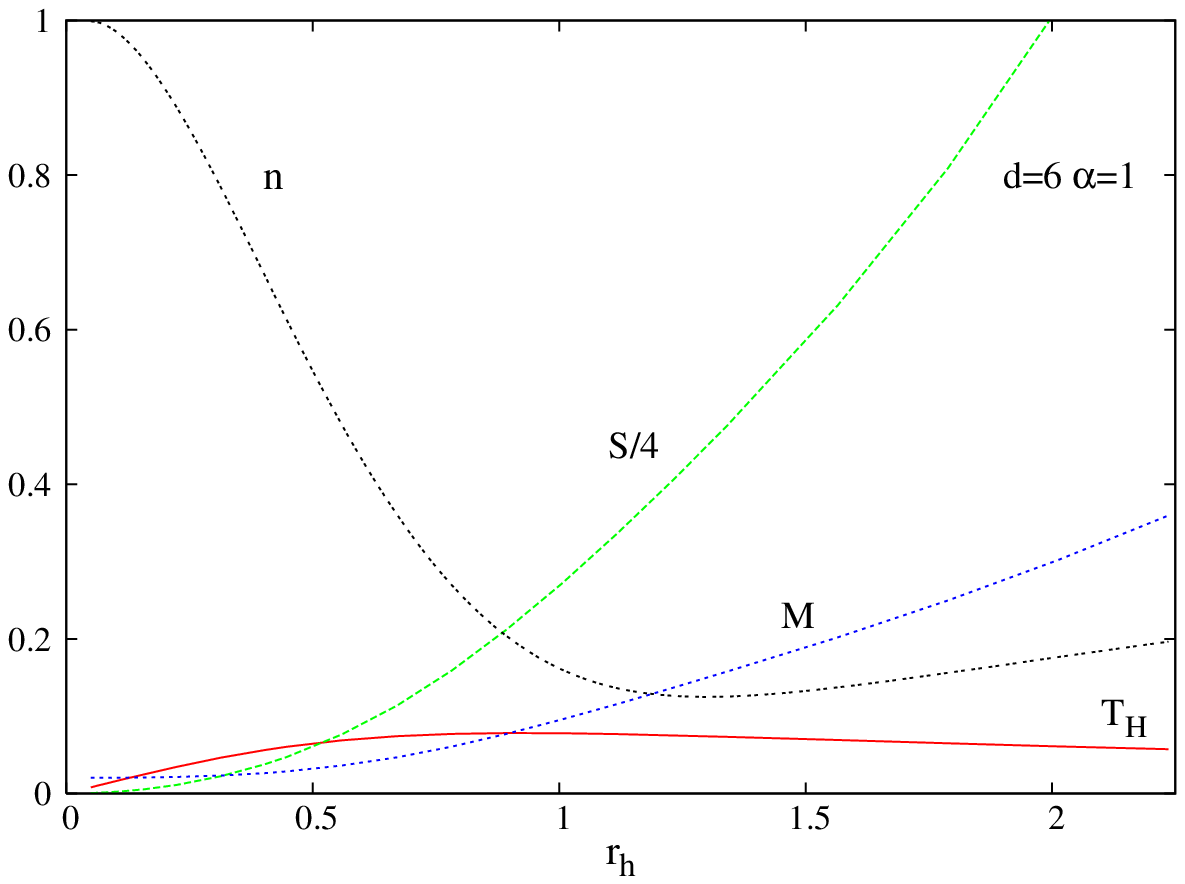}}
\hspace{5mm}%
        \resizebox{8cm}{6cm}{\includegraphics{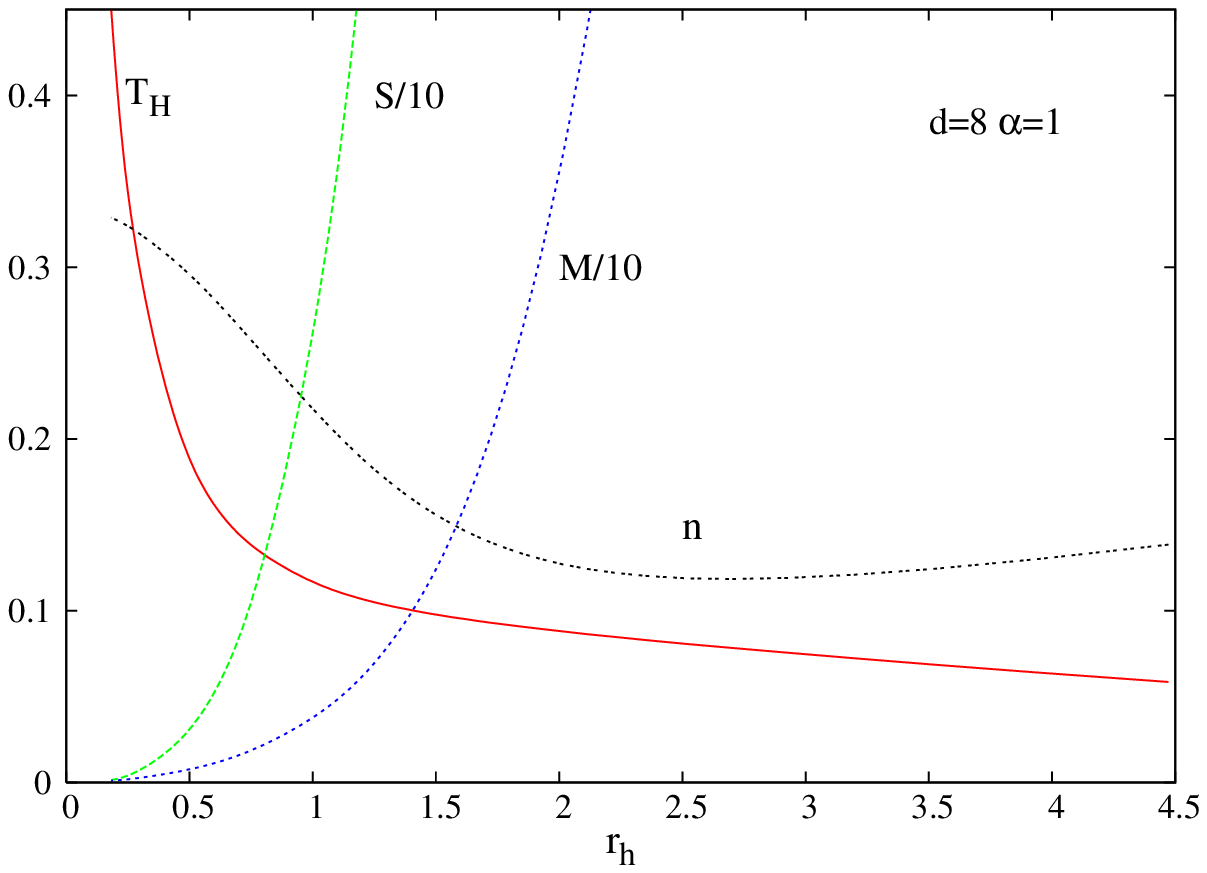}}	
\hss}

\caption{
{\small
The $M$, the relative tension $n$, the Hawking temperature $T_H$ and the entropy $S$ 
of the $d=6,8$  uniform string solutions
are shown 
 as a function of the event horizon radius $r_h$, for a fixed value $\alpha=1$ of the GB coupling constant.
 For the data here and in Figure 8 we have set $V_{d-3}L/G=1$.}
}
\label{var-rh-therm}
\end{figure}

 In the case $d=8$, our classification of the fixed points \eqref{fpd8} 
 shows that $b(r)$ becomes singular at the origin for all the solutions except for 
 the flat spacetime background. As a consequence, the 'Kaluza-Klein fixed point' 
 appears as the most likely configuration into which the black string 
 can be attracted for $r \in ]0,\infty]$ in the limit $r_h \to 0$.

We close this part by noticing that,
in the presence of curvature-squared terms,
the modified Einstein equations leads to an effective stress tensor that involves the gravitational field
$ G_{\mu\nu}=-\alpha H_{\mu\nu}\equiv T_{\mu\nu}.$
Therefore, from some point of view, the quantity $\alpha H_{t}^t(=-G_t^t)$ corresponds to a
local  `effective energy density'.
However, this effective stress tensor, 
thought of as a kind of matter distribution,
in principle may violate the weak energy 
condition. This property has been noticed in \cite{Kleihaus:2009dm}
for $d=5$ black string solutions, being employed there to explain some properties
of the black rings in EGB theory.
As one can see in Figure \ref{Htt}, this feature of the five dimensional solutions is also valid for $d>5$
black strings with $\alpha>0$.

\subsubsection{Thermodynamical aspects}

The various thermodynamical quantities were extracted from the numerical output. 
Our results satisfy the Smarr relation \eqref{smarrform} with a precision of $10^{-3}$. 
We have also integrated the first law of thermodynamic \eqref{firstlaw} to crosscheck 
the mass obtained in this way for fixed values of $L$ with the mass (\ref{2}) computed from the 
asymptotic quantities $c_t$ and $c_z$. 
These two computations of the mass are in excellent agreement, which provides 
a strong test of the accuracy of our numerical results.

We start by presenting here the various thermodynamical quantities as functions of 
the event horizon radius $r_h$, for a fixed value of the GB coupling $\alpha$. 
Figure \ref{var-rh-therm} (left) presents the entropy, Hawking temperature,
  mass and relative tension  
for $d=6$ solutions. Although the data there has $\alpha=1$, this is the generic picture, since
the results for other finite (and nonvanishing) values of $\alpha$ can be reconstructed by using the 
scaling (\ref{scal-new}).

\begin{figure}[ht]
\hbox to\linewidth{\hss%
	\resizebox{8cm}{6cm}{\includegraphics{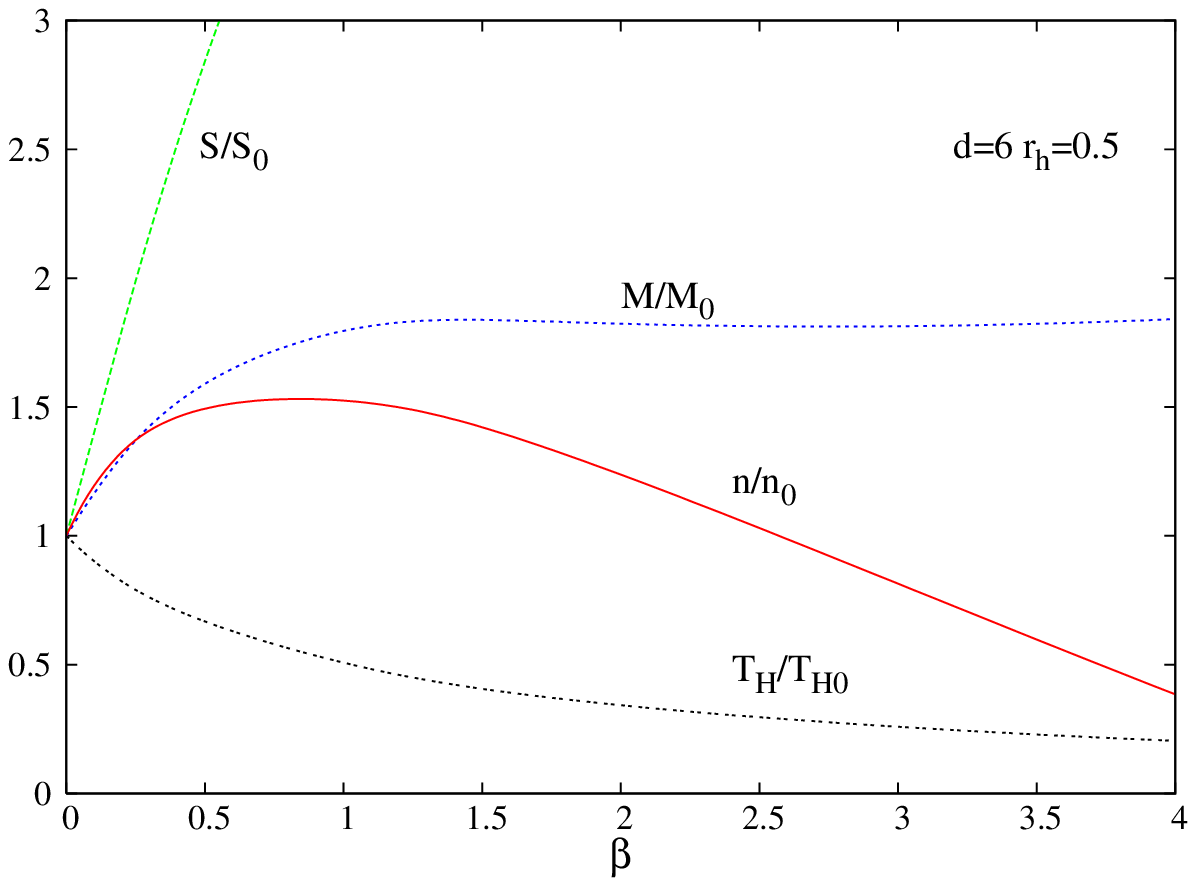}}
\hspace{5mm}%
        \resizebox{8cm}{6cm}{\includegraphics{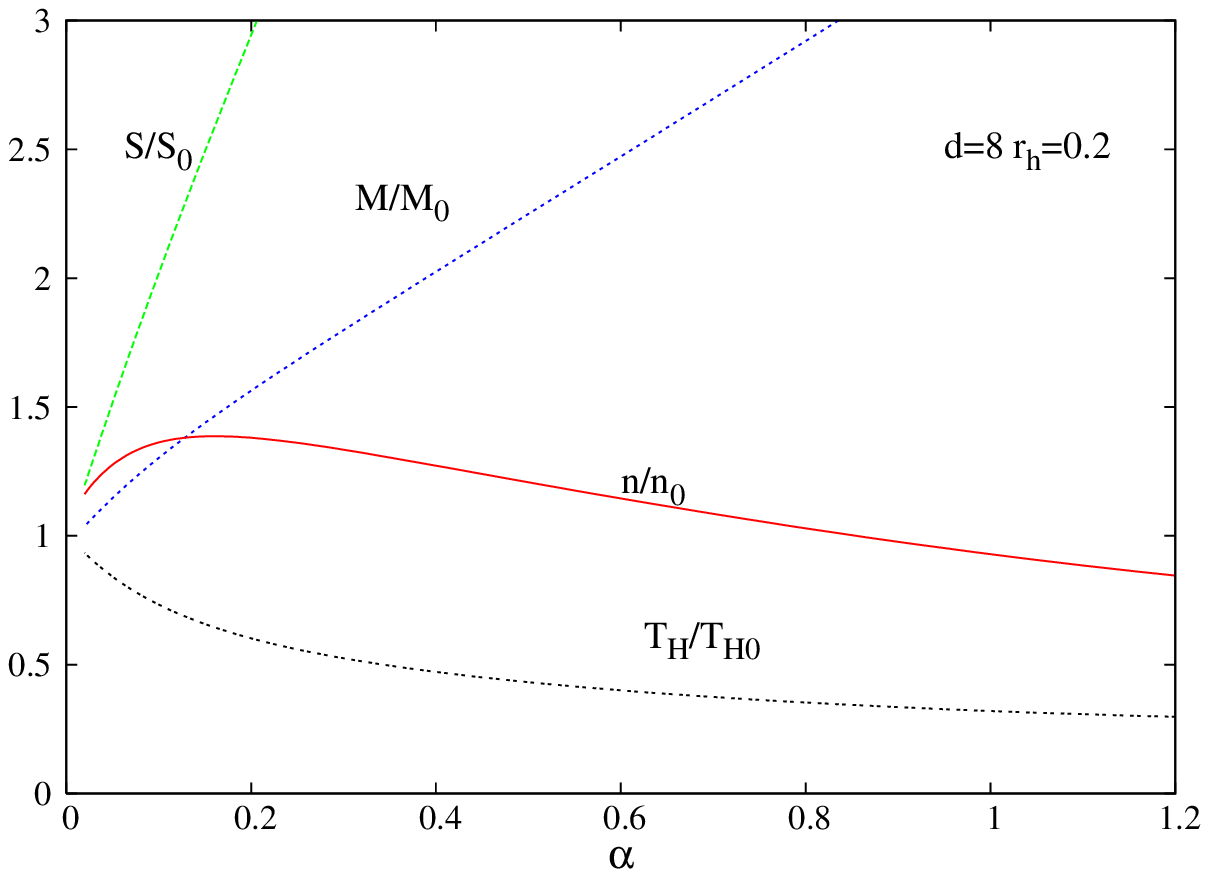}}	
\hss}

\caption{
{\small
The $M$, the relative tension $n$, the Hawking temperature $T_H$ and the entropy $S$ 
of the $d=6,8$  uniform string solutions
are shown 
 in units of the Einstein gravity uniform string solution
(denoted by $M_0$, $n_0$, $T_0$, and $S_0$) as functions of
 the dimensionless parameter $\beta=\alpha/r_h^2$.}
}
\label{fig43}
\end{figure}

It can be observed that the temperature reaches a maximum
at a particular value of $r_h$
 while the entropy increases monotonically with $r_h$; 
 this implies the existence for $d=6$ of two phases of EGB black strings, 
 one thermodynamically stable, the other thermodynamically unstable. 
 (We will return to 
 this property in the next subsection.) 
We note also that the mass and tension do not vanish in the limit  $r_h\rightarrow0$.

Figure \ref{var-rh-therm} (right) presents the same quantities for $d=8$ and $\alpha=1$. 
In this case, there is only one thermodynamically unstable phase since the temperature is always decreasing 
with $r_h$, while the entropy is always increasing. 
For $d=8$, the mass and tension approach zero as $r_h\rightarrow0$, 
confirming  our interpretation that the solution approaches  the flat 
spacetime (\ref{KK-metric}) in this limit.

In Figure 6 we illustrate the dependence of various  thermodynamical quantites  on the parameter
$\beta=\alpha/r_h^2$. There we have kept fixed the value of the event horizon radius and thus that data
show how the solutions depend on the GB coupling constant $\alpha$. 
The quantities there are given
 in units of the corresponding Einstein gravity uniform string solutions.
 An interesting feature here is the different behaviour of the  solutions' mass.
 For $d=8$, this quantity increases almost linearly with $\beta$,
 while for $d=6$ it approaches a constant for large $\beta$.

\section{Stability of the EGB uniform black strings}
As shown by Gregory and Laflamme, for a given circle size, the Einstein gravity uniform 
black strings below a critical mass are  linearly unstable.
That is, at the critical point, there is a static mode breaking the
translational invariance on the circle.
This instability can be understood
heuristically by using an entropy argument: for a
black string horizon with $S^{d-3}\times S^1$ topology, there exist
a length $L$ above which it becomes
entropically favourable for the mass
to localize in a black hole with spherical topology.
This connection between classical and thermodynamical
stability has been made more precise with a
conjecture formulated by Gubser and Mitra (GM) \cite{Gubser:2000ec}.
Their conjecture is that for systems with
translational symmetry and infinite extent, a GL
instability arises precisely when the system becomes
thermodynamical unstable.

Here we propose to find how  a GB term will affect these results.
The case $d=5$ has been considered in \cite{Suranyi:2008wc}. 
The results there show that GL instability persists for all allowed values of the
GB coupling constant.

\subsection{Perturbations of the uniform metric }

In addressing the question of stability of the black string solutions presented above, 
we have found most convenient to use 
 the approach of Ref. \cite{Gubser:2001ac} and to consider in particular a NUBS metric ansatz 
\be
ds^2 =-b(r) e^{2A(r,z)}dt^2 + e^{2B(r,z)}\left( \frac{dr^2}{f(r)} + a(r)dz^2 \right) + r^2e^{2C(r,z)} d\Omega_{d-3}^2.
\label{nuans}
\ee
where the non-uniformity is 
set in through the functions $A,B,C$. The uniform solutions discussed in Section 3 are recovered for $A=B=C=0$.

Then,  following \cite{Gregory:1993vy} we perform an expansion of the functions $A,B,C$ in terms 
of a small parameter $\epsilon$ and consider a Fourier series in the $z$ coordinate. In the leading order, we assume:
\be
X(r,z) = \epsilon X_1(r) \cos(k z) + \Ord{\epsilon}{2},
\label{Xseries}
\ee
$X$ denoting generically $A,B,C$ and $k$ being the critical wavenumber corresponding to a static perturbation~:
$k= 2 \pi/L$.

We then substitute the form \eqref{nuans} in the general EGB equations (\ref{eqs}) and  
expand $A,B,C$ according to \eqref{Xseries}.
In the next step, we consider the linear order terms there in the infinitesimal parameter $\epsilon$. 
The system obtained in this way is the relevant one for addressing the stability problem.
Similarly to the pure Einstein case (for both flat \cite{Gubser:2001ac} and AdS \cite{Brihaye:2007ju} black strings),
 the $t-r$ component of the EGB equations allow to eliminate the field  $B_1$ in favor of the other fields
 and to reduce the problem to a system of two differential equations for $A_1$ and $C_1$.
 These equations are on the form:
 \be
A_1''= \psi_1 A_1 + \psi_2 A_1' + \psi_3 C_1 + \psi_4 C_1',~~
C_1'' = \varphi_1 A_1 + \varphi_2 A_1' + \varphi_3 C_1 + \varphi_4 C_1',
\label{eqF1}
\ee 
 where the expressions of $\psi_i,\varphi_i$ are quite involved, 
 so we prefer not writing them here\footnote{
We have verified that  the Einstein gravity equations of \cite{Gubser:2001ac} are recovered in the limit $\alpha\rightarrow0$. 
}
 (they depend on the metric functions $a,b,f$ and their first derivatives).
 
Since we want to describe \emph{perturbations}, the functions $A_1,\ C_1$ should obey the following 
 conditions asymptotically:
\be
A_1(\infty) = 0,\ C_1(\infty)=0.
\ee
The requirement that the perturbations obeying the linearized equations should be regular at the horizon 
leads to specific relations to be satisfied by $A_1(r_h)$, $C_1(r_h)$ and their derivatives. 
 The final expressions are really long due to the GB terms
 and we do not write them explicitely. We just mention that they are of the form
\bea
\label{reg}
A_1'(r_h) &=& \mathcal A(r_h,d,\alpha;a_0,b_1;A_1(r_h),C_1(r_h)),
\\
C_1'(r_h) &=& \mathcal C(r_h,d,\alpha;a_0,b_1;A_1(r_h),C_1(r_h)),
\nonumber
\eea
where $a_0,b_1$ appear in the UBS expansion (\ref{exphor}).
Finally, the arbitrary normalisation in the linear system is fixed by 
\be
C_1(r_h)=1.
\ee
This leads to five boundary conditions which can be accomodated only if the parameter $k^2$ assumes very specific values.
The system of equations under consideration 
is therefore similar to a spectral problem. 
The sign of $k^2$ determines the stability:
if $k^2$ is positive, there exists a GL unstable regime  \cite{Gregory:1993vy} for values of the wavenumber, say $k$, 
smaller than the critical wave number.  If $k^2$ is negative, the system is stable.

%
%
%
\subsection{Numerical procedure and results}
 The (linear) stability  equations have been solved by using both the solver COLSYS \cite{colsys}
 (which employs a Newton-Raphson method) and
 a two parameters shooting method. 
 This has allowed to cross-check the results and 
 to obtain the pattern of stability  as accurately as possible.   
 The shooting parameters can be chosen as the critical wavenumber $k^2$ and the value of the function $A_1$ at the horizon. 
 Together with the regularity conditions \eqref{reg}, this leads to a Cauchy problem, suitable for a shooting method.

\begin{figure}[ht]
\hbox to\linewidth{\hss%
	\resizebox{10cm}{8cm}{\includegraphics{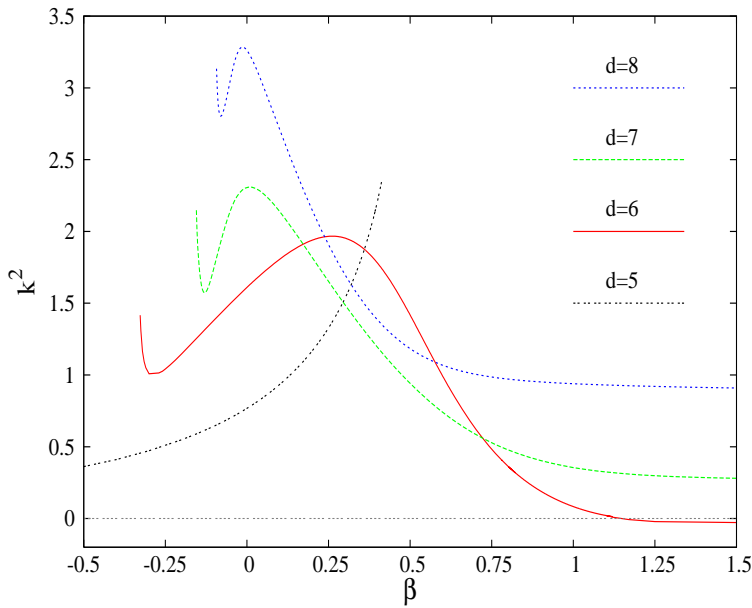}}
\hss}

\caption{
{\small
The critical wavenumber is shown as a function of the dimensionless parameter $\beta=\alpha/r_h^2$ for $d=5,6,7,8$. 
The critical wavenumber diverges for $\beta=-1/3$ in $d=6$ 
(resp. $\beta=-1/6$ for $d=7$ and $\beta=-1/10$ for $d=8$) 
and asymptotes to a constant for large values of $\beta$. 
For $d=5$, one finds  $-1<\beta<1/2$, the complete picture there being presented in 
Ref. \cite{Suranyi:2008wc}.
}
}
\label{k2-vs-alpha}
\end{figure}

The case $d=5$ has been considered in \cite{Suranyi:2008wc}. 
The value of the dimensionless GB coupling constant $\beta$
 ranges\footnote{
 Note that the definition of the rescaled Gauss-Bonnet coupling $\beta$
in this paper is different than the one $\tilde \beta$ used in \cite{Suranyi:2008wc},
with $\beta=(1-\tilde \beta)/2$.
} 
here in the interval $[-1,1/2]$. 
The results in \cite{Suranyi:2008wc} show that the critical wavenumber $k^2$
is always positive and
 remains finite  for $\beta$ approaching the  maximal value $\beta=1/2$, and that $k^2$
 diverges to infinity  in the limit $\beta \to -1$.  
Our numerical results seem to confirm this behaviour, although we have found difficult
to approach the limits
of the $\beta-$interval (see Figure  \ref{k2-vs-alpha}).
However, one can conclude that all $d=5$ EGB
 black strings possess a GL instability. 
 
A main difference between the  $d>5$ and the five dimensional case
is that the solutions exist for large positive values of $\beta$.
Our numerical results show that
the  critical wavenumber $k^2$ goes to a constant for large values of $\beta$, as 
conjectered in \cite{Suranyi:2008wc}.
However, we have found that this constant is negative for $d=6$.
Therefore, 
the EGB black strings are stable for large values of $\beta$ in six dimensions. 
In other words, for a given event horizon radius, 
one can stabilize an EGB black string by taking
a large enough value of the GB coupling constant. 

In the cases $d>6$, the square critical wavenumber takes strictly positive values, $i.e.$ the solutions remain unstable
for any value of $\alpha$.
  These results are illustrated in Figure \ref{k2-vs-alpha} for $d=6,7,8$ and we expect a similar behaviour 
 to hold for $d>8$. 
That plot also shows that the value $k^2_{\infty} \equiv \lim_{\beta \to \infty} k^2$ increases with the number
of dimensions of space-time.
Also, as seen in Figure \ref{k2-vs-alpha},
the critical length $L_c$ of the extra-direction shows a complicated dependence on $\alpha$.
However, for large enough $\alpha$ and $d>6$, $L_c$ increases with the GB coupling constant, approaching asymptotically
a constant value.
 
When the minimal value of $\beta$
(\ref{lbound}) is approached, we note that $k^2$ increases and likely diverges to infinity 
(note, however that the numerics
becomes more difficult as $\beta\to \beta_c$).   
This behaviour, which also holds for $d=5$ (see the Figure 3 in \cite{Suranyi:2008wc} and the footnote 11),
appears also for the $d>5$ data exhibited in Figure \ref{k2-vs-alpha}.

One should notice that the perturbation (\ref{Xseries})
can be generalized to higher order, by taking into account the non-linear back reaction.
For Einstein gravity black strings, the study of the perturbative equations in second order revealed
the appearance of a critical dimension, above which the perturbative non-uniform black strings
are less massive than the marginally stable uniform black string \cite{Sorkin:2004qq}. 
It would therefore be
interesting to solve the perturbative equations to second order also in the presence of a GB term,
the only obstacle we can see at this moment 
being the tremendous complexity of the 
equations.

%
%
\begin{figure}[ht]
\hbox to\linewidth{\hss%
	\resizebox{8cm}{6cm}{\includegraphics{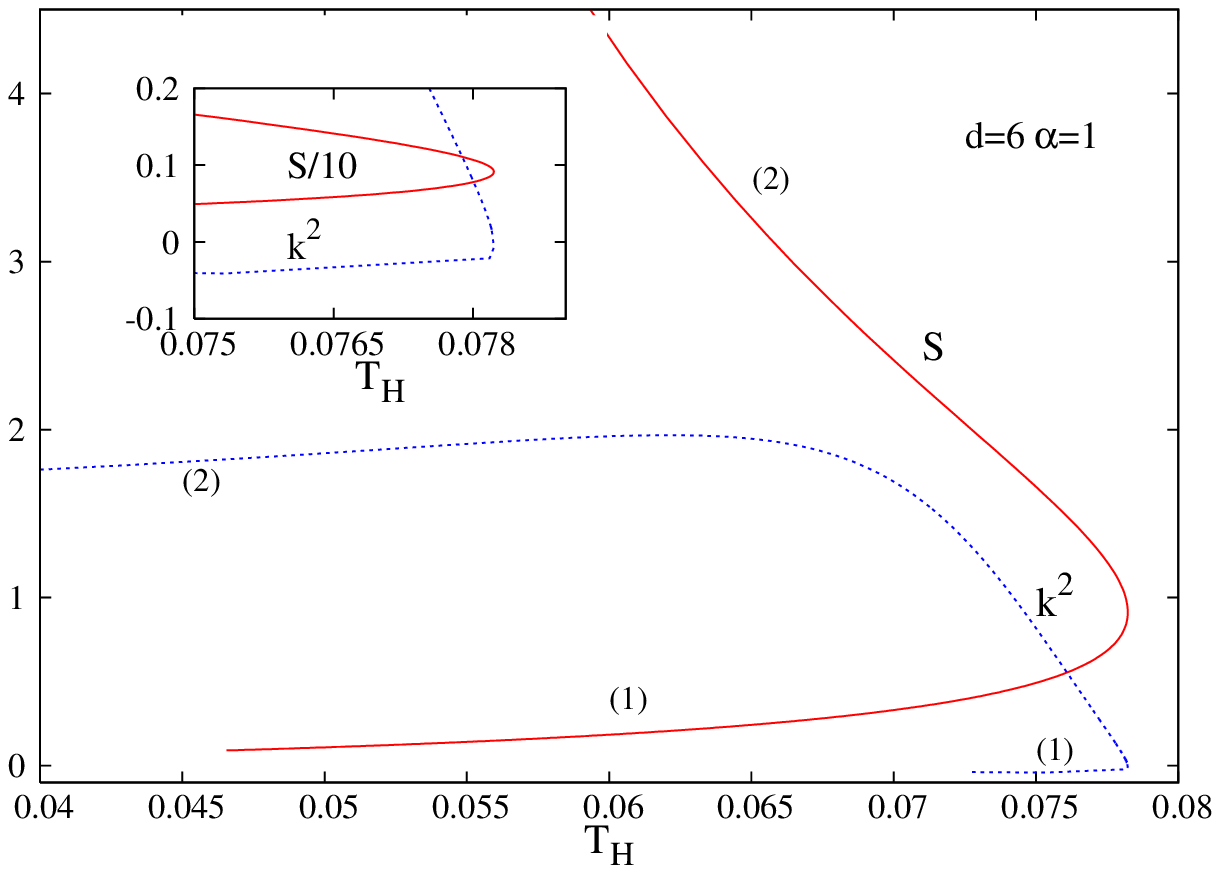}}
\hspace{5mm}%
        \resizebox{8cm}{6cm}{\includegraphics{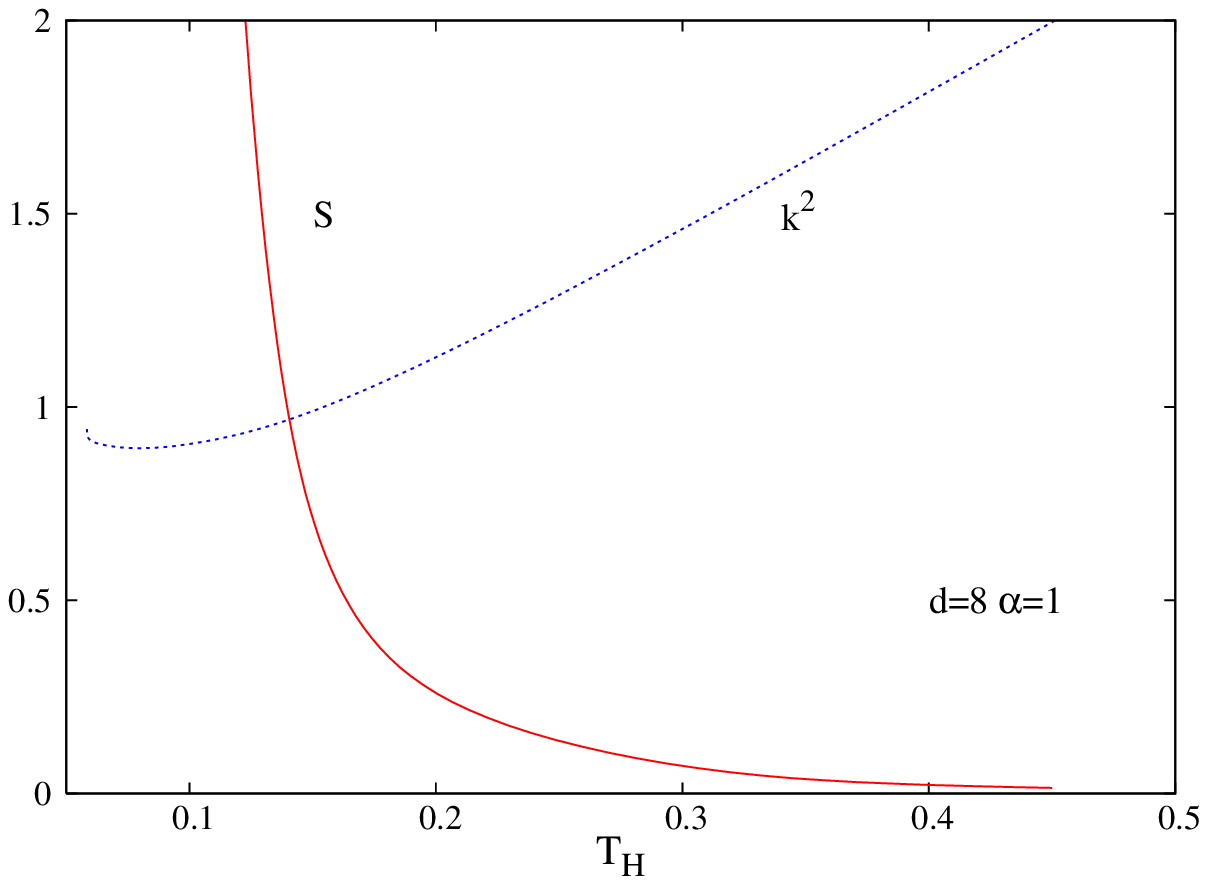}}	
\hss}

\caption{
{\small
The entropy and the square critical wavenumber of $d=6,8$ EGB uniform black strins
are shown as a function of the Hawking temperature.
The $d=8$ solutions have only one thermodynamically unstable
phase.
}
}
\label{GM}
\end{figure}
%
\subsection{Correlated stability conjecture}

We have seen that the family of EGB black strings present  a dynamically stable phase in $d=6$. 
However, for $d>6$, we didn't find such a phase. In the following, we discuss this feature
in relation with the  thermodynamical stability of the EGB strings in $d=6$ and $d>6$. 
In the absence of matter fields  or rotation,
the study of the thermodynamical stability of a black string is related to the sign of the heat capacity
\be
C_p = T_H\left(\frac{\partial S}{\partial T_H}\right)\bigg|_{L}.
\ee
Solutions with  $C_p$ positive (respectively negative)  are stable (respectively unstable). 
From the GM correlated stability conjecture \cite{Gubser:2000ec}, 
it is expected that  both a thermodynamically stable and unstable phases exist for $d=6$ and that
$C_p$ is always negative for $d>6$.

We have focused the analysis on $d=6$ and $d=8$, hoping that 
 the dimension $d=8$ 
catches the main features for other values of $d>6$.

Figure \ref{GM} (left) presents the entropy $S$ and the square critical wavenumber
$k^2$ as  functions of the Hawking temperature for $d=6$. 
One can see that the solutions with $k^2<0$
have also a positive specific heat (this branch has the index $1$ in that plot), 
while
the entropy decreases with $T_H$ along the branch with $k^2>0$ 
(index $2$ in the Figure \ref{GM}).
In agreement with the GM conjecture,  
the critical temperature where $k^2$ crosses zero coincides 
(within the numerical accuracy) with
the temperature at the turning point of the UBS branches in a $(S-T_H)$ diagram. 
Therefore the correlated stability conjecture
holds for the family of solutions under consideration.

For higher values of the number of dimensions, there is no 
thermodynamically stable phase, as shown by Figure \ref{GM} for $d=8$.
As a result, the critical wavenumber $k^2$ takes strictly positive values.
 
The different picture one finds for $d=6$ can be understood heuristically by taking into account that
the black strings should 
inherit some of 
the features of the BHs in $d-1$ dimensions.
However, as discussed in the Appendix A, the $d=5$ Schwarzschild black hole
in EGB theory has a positive specific heat for large enough $\beta$.
For $d>5$, the spherically symmetric black holes are thermodynamically unstable,
and thus also the corresponding $d>6$ uniform black strings.

\section{$d=6$ non-uniform EGB black string}

As realized in \cite{Gregory:1993vy}, \cite{Gubser:2001ac},
the GL classical instability of the UBSs implies the existence 
of a new branch of solutions.
The non-uniform black strings are known in Einstein gravity
for all dimensions from five to eleven  \cite{Wiseman:2002zc}, \cite{Kleihaus:2006ee},
\cite{Sorkin:2006wp}.
These configurations are non-uniformly distributed
along the $z$-direction and are continuously connected to the uniform
strings, possesing the same horizon topology.
However, for sufficient
non-uniformity, the local geometry about the minimal horizon sphere (the �waist�)  
tends to a cone metric.
The numerical results presented in \cite{Kudoh:2004hs}, \cite{Kleihaus:2006ee} 
suggest that, at least for $d=5,6$, the non-uniform
string branch merges with the caged black hole branch at a topology changing transition.

Although all these solutions should present generalizations  with a GB term, we
restrict our study to the simplest case $d=6$. 
This is also the most studied in Einstein gravity\footnote{The $d=6$ NUBS  have been 
constructed independently, by several authors \cite{Wiseman:2002zc}, \cite{Kleihaus:2006ee},
\cite{Sorkin:2006wp}, with rather different methods.},
the numeric being easier for this spacetime dimension. 
Here we do not aim on a systematic study of these solutions, which would be a difficult taks.
Instead, we want to establish the existence of EGB non-uniform solutions 
and to find their basic properties.
Moreover, we have restricted our analysis to solutions with a positive value of 
the GB coupling constant (although we have verified the existence of solutions with $\alpha<0$
as well).

\subsection{The ansatz and boundary conditions}

The non-uniform black string solutions presented in this paper 
are found within a metric ansatz  employed also for the study of
$d=6$ NUBS in pure Einstein gravity\footnote{Although the metric ansatz (\ref{nuans}) is useful in the perturbation theory,
we could not employ it for the study of non-uniform solutions.}:
\begin{eqnarray} 
\label{metric}
ds^2=-e^{2A(r,z)}\frac{r^2}{r^2+r_0^2}dt^2+e^{2B(r,z)}\left( dr^2 +dz^2\right)+e^{2C(r,z)}(r^2+r_0^2) d\Omega_3^2, 
\end{eqnarray}
with $r_0$ a positive constant.
The event horizon resides at $r=0$ where $g_{tt}=0$
and $A,B,C$ stay  finite.

The equations for the functions $A,~B$ 
and $C$ are found by using a suitable combination of the EGB equations,
$E_t^t =0,~E_r^r+E_z^z =0$
and  $E_{\varphi}^{\varphi} =0$ (with $\varphi$ an angle on $\Omega_3$)
 which diagonalizes the Einstein tensor w.r.t. $\nabla^2 A$, $\nabla^2 B$, $\nabla^2 C$ 
 (where $\nabla^2=\partial_{rr}+\partial_{zz}$). 
Due to the GB contribution, these equations 
are much more complicated than in the case of Einstein gravity (with around 200 terms each equation)
and we shall not present them here.

The remaining equations $E_z^r =0,~E_r^r-E_z^z  =0$
yield two constraints. 
Following \cite{Wiseman:2002zc}, we note that
setting $E^t_t =E^{\varphi}_{\varphi} =E^r_r+E^z_z=0$
in $\nabla_\mu E^{\mu r} =0$ and $\nabla_\mu E^{\mu z}=0$, 
we obtain the Cauchy-Riemann relations
\begin{eqnarray}
\partial_r {\cal P}_1  +
\partial_z {\cal P}_2  
= 0 ,~~
 \partial_z {\cal P}_1  
-\partial_r {\cal P}_2
~= 0 ,~~~{~~~} 
\end{eqnarray}
with ${\cal P}_1=\sqrt{-g} E^r_z$, ${\cal P}_2=\sqrt{-g}(E^r_r-E^z_z)/2$.
Therefore the weighted constraints still satisfy Laplace equations, and the constraints 
are fulfilled, when one of them is satisfied on the boundary and the other 
at a single point
\cite{Wiseman:2002zc}.

Utilizing the reflection symmetry of the non-uniform black strings 
w.r.t.~$z=L/2$,
the solutions are constructed subject to the following set of 
boundary conditions
\begin{eqnarray}
\nonumber
&&A\big|_{  r=\infty}=B\big|_{ r=\infty}=C\big|_{ r=\infty}=0,
\\
\label{bc2} 
&&A\big|_{  r=0}-B\big|_{  r=0}=d_0,~\partial_{  r} 
A\big|_{ r=0}=\partial_{  r} C\big|_{  r=0}=0,
\\
\nonumber
&&\partial_z A\big|_{z=0,L/2}=\partial_z B\big|_{z=0,L/2}
=\partial_z C\big|_{z=0,L/2}=0,
\end{eqnarray}
where the constant $d_0$ is related to the Hawking 
temperature of the solutions.
Regularity further requires that the condition 
$\partial_{  r} B\big|_{  r=0}=0$ holds for the solutions.
The boundary conditions guarantee, that the constraints are satisfied,
since $\sqrt{-g} E^r_z=0$ everywhere on the boundary,
and $ \sqrt{-g}( E^r_r-E^z_z ) =0$ on the horizon.

One should note that, different from the
Einstein gravity case, the functions $A,B,C$ have a nontrivial shape for the uniform string,
which has also $d_0\neq 0$.

The Hawking temperature and entropy of the black string solutions are given by
\begin{eqnarray}
\label{temp} 
&T_H=\frac{e^{A_0-B_0}}{2 \pi r_0},~
S= \frac{1}{4G} V_{3}r_0^{3} \int_0^L dz e^{B_0+3C_0}
\bigg[1+\frac{3\alpha}{r_0^2}
\left( e^{-2C_0}+e^{-2B_0}r_0^2(B_{0,z}C_{0,z}-2C_{0,z}^2-C_{0,zz})\right)\bigg],~~{~~}
\end{eqnarray}
where  $A_0(z),B_0(z),C_0(z)$ are the values of the metric functions on the event horizon $r=r_0$.
The mass and tension of a NUBS solution are still given by 
(\ref{2}), where $c_t$ and $c_z$ are the asymptotic coefficients in the 
metric functions $g_{tt}=-e^{2A(r,z)}{r^2}/{(r^2+r_0^2)}$ and
$g_{zz}=e^{2B(r,z)}$, respectively.

A measure of the deformation of the solutions is given by the
non-uniformity parameter $\lambda$ \cite{Gubser:2001ac}
\begin{equation} 
\lambda = \frac{1}{2} \left( \frac{{\cal R}_{\rm max}}{{\cal R}_{\rm min}}
 -1 \right)
, \label{lambda} \end{equation}
where ${\cal R}_{\rm max}$ and ${\cal R}_{\rm min}$
represent the maximum radius of the three-sphere on the horizon and
the minimum radius, being the radius of the `waist'.
Thus the uniform black strings have $\lambda=0$,
while the conjectured horizon topology
changing transition should be approached
for $\lambda \rightarrow \infty$ 
\cite{Kol:2003ja,Wiseman:2002ti}.

\subsection{Numerical methods}
While, as we shall see, their basic physical properties
are similar to those of the Einstein gravity counterparts, 
the construction of  NUBS in EGB theory is a more difficult task.
This is due, perhaps, to
the much more complicated form of the equations.

To obtain these solutions,
we solve the set of three coupled non-linear
elliptic partial differential equations numerically,
subject to the  boundary conditions (\ref{bc2}).
The approach here is similar to that employed in 
\cite{Kleihaus:2006ee} to construct $d=5,6$
NUBS in Einstein gravity.
We employ a compactified radial coordinate
$\bar r = r/(1+ r)  $
which
maps spatial infinity to the finite value $\bar r=1$.
The numerical calculations are based on the Newton-Raphson method
and are performed with help of the program FIDISOL/CADSOL \cite{schoen},
which provides also an error estimate for each unknown function.
For the solutions in this work,
the typical  numerical error 
for the functions is estimated to be lower than $10^{-3}$. 
 
The equations are discretized on a non-equidistant grid in
$\bar r$ and $z$.
Typical grids used have sizes $90 \times 50$,
covering the integration region
$0\leq \bar r \leq 1$ and $0\leq z \leq L/2$. 
The horizon coordinate $r_0$ and the asymptotic length $L$ of the compact
direction enter the equations of motion as parameters.
The results presented are mainly obtained with the parameter choice
$r_0=1$, while the value of $L$ is $\alpha$-dependent. 
Also, due to numerical difficulties, our solutions have 
rather small values of the GB coupling constant $\alpha<0.05$.
New features may appear for larger values, in particular for $\alpha/r_0^2$ close to one, a region which we could not explore
with our methods.

The most error-prone part of our numerical computation is to extract
the asymptotic coefficients
$c_t,~c_z$.
Different from the Einstein gravity solutions, 
there is no Smarr relation\footnote{This is not a surprise,
given the presence of an extra-length
scale fixed by $\alpha$. We recall that no Smarr relation can be found
even for  the asymptotically flat black hole solutions
 in EGB theory.} which can be used to verify the global accuracy of the results.
The existence of the Smarr relation (\ref{smarrform}) for UBSs was 
a consequence of the existence of a Killing vector $\partial/\partial z$, which implies the
relation (\ref{totder2}).
In fact, we have noticed that (\ref{smarrform}) does not hold for NUBSs, 
the deviation increasing with $\lambda$.
However, the Hawking temperature, the entropy and the non-uniformity parameter $\lambda$
are measured directly from the metric and thus from a numerical
point of view they are 'cleaner' than the mass and tension.

\subsection{Properties of the solutions}
Similar to the case of Einstein gravity,
a branch of non-uniform strings emerges smoothly from
the uniform string branch at the GL critical point,
where its stability changes.
For a given value of $\alpha$, keeping the horizon coordinate $r_0$ and the 
corrresponding asymptotic
length $L$ of the compact direction fixed,
the solutions depend on the parameter $d_0$ only\footnote{We recall that,
within the ansatz (\ref{metric}), 
the UBSs have $d_0\neq 0$.}, which enters the Eq.~(\ref{bc2}).
Increasing this parameter, the non-uniform strings
become more and more deformed, 
as quantified by the non-uniformity
parameter $\lambda$. 

In Figure \ref{NUBS-3d} we exhibit the metric functions $e^{2A}$, $e^{2B}$, and
$e^{2C}$ for two typical solutions with different values of the GB
coupling constant $\alpha$
and the same value of $T_H$, $r_0$ and $L$.
One can see that
the shape of these functions is similar to that exhibited in the literature
for $\alpha=0$.
Moreover, one finds the same plots
for the spatial embedding of the horizon into 3-dimensional space.

The $A,B,C$ functions exhibit extrema on the symmetry axis $z=0$
at the horizon. 
As $\lambda$ increases, the extrema increase in height and become
increasingly sharp. 
As $\lambda \to \infty$, one expects that the 'waist' of the NUBS would shrink to zero, locally
forming a cone geometry, which should connect to the EGB caged black hole branch.
Unfortunately, the numerical accuracy  deteriorates 
for large values of $\lambda$.

\newpage
\setlength{\unitlength}{1cm}
\begin{picture}(15,18)
\put(-1,0){\epsfig{file=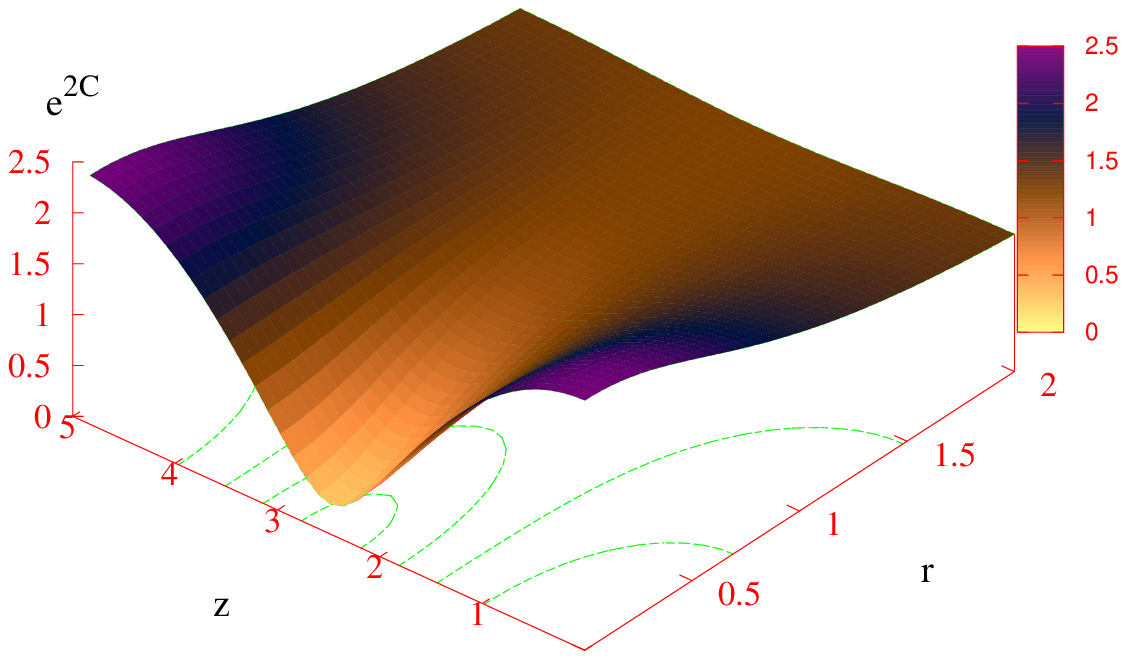,width=8cm}}
\put(7,0){\epsfig{file=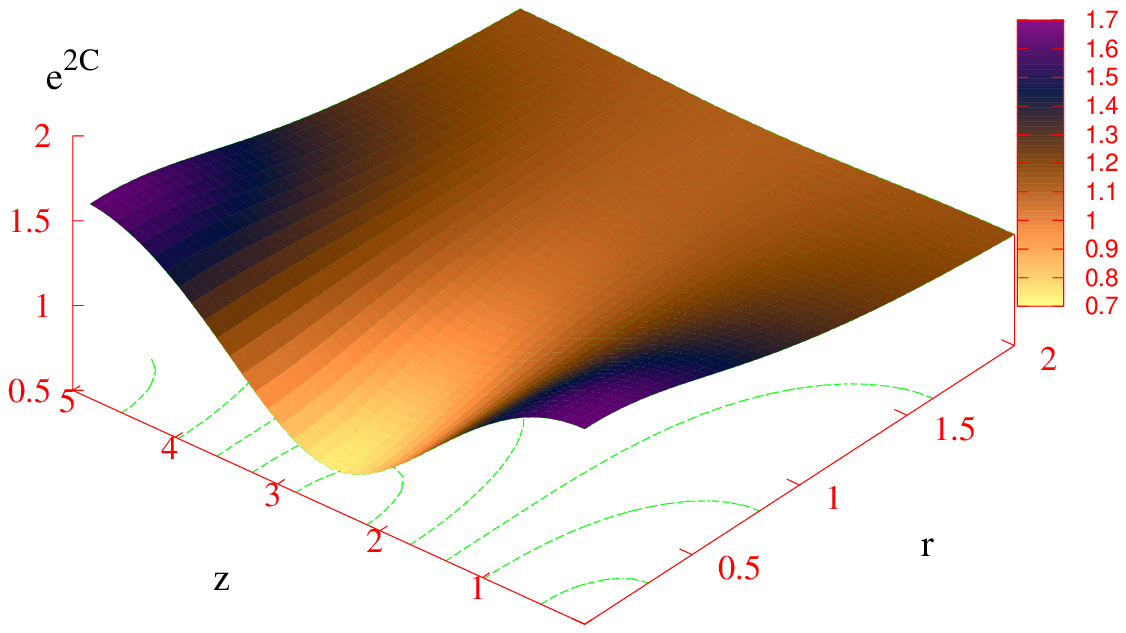,width=8cm}}
\put(-1,6){\epsfig{file=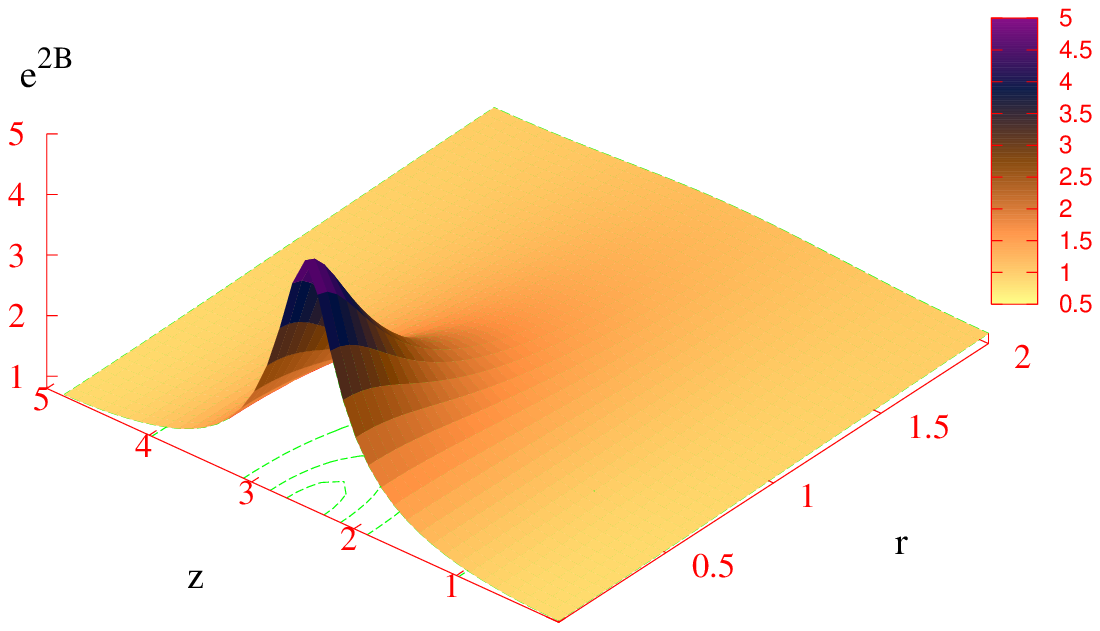,width=8cm}}
\put(7,6){\epsfig{file=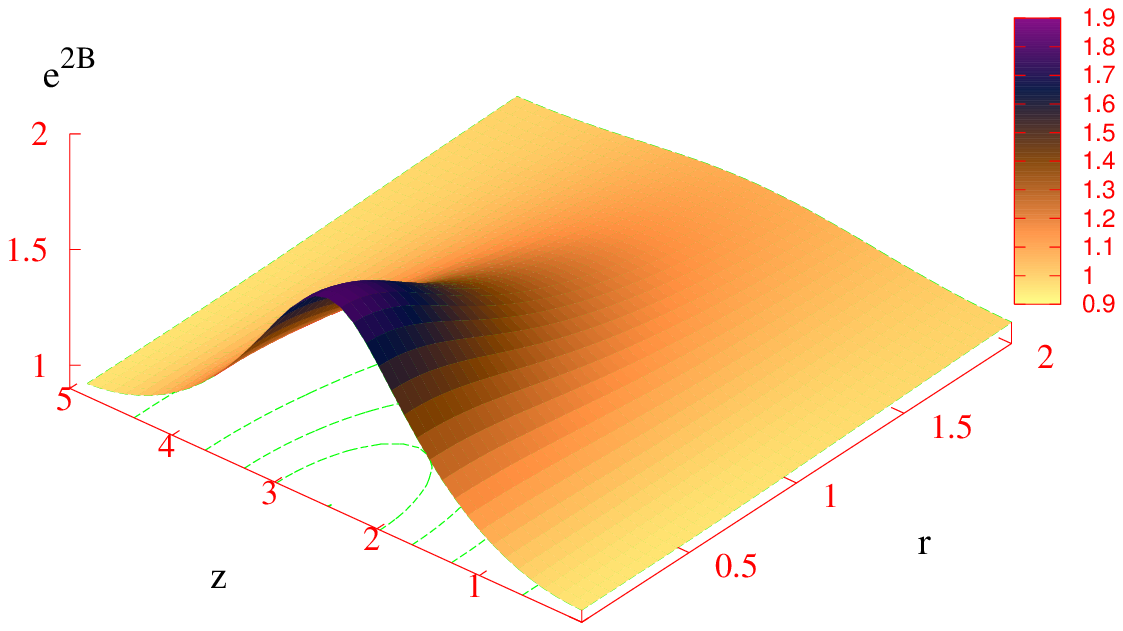,width=8cm}}
\put(-1,12){\epsfig{file=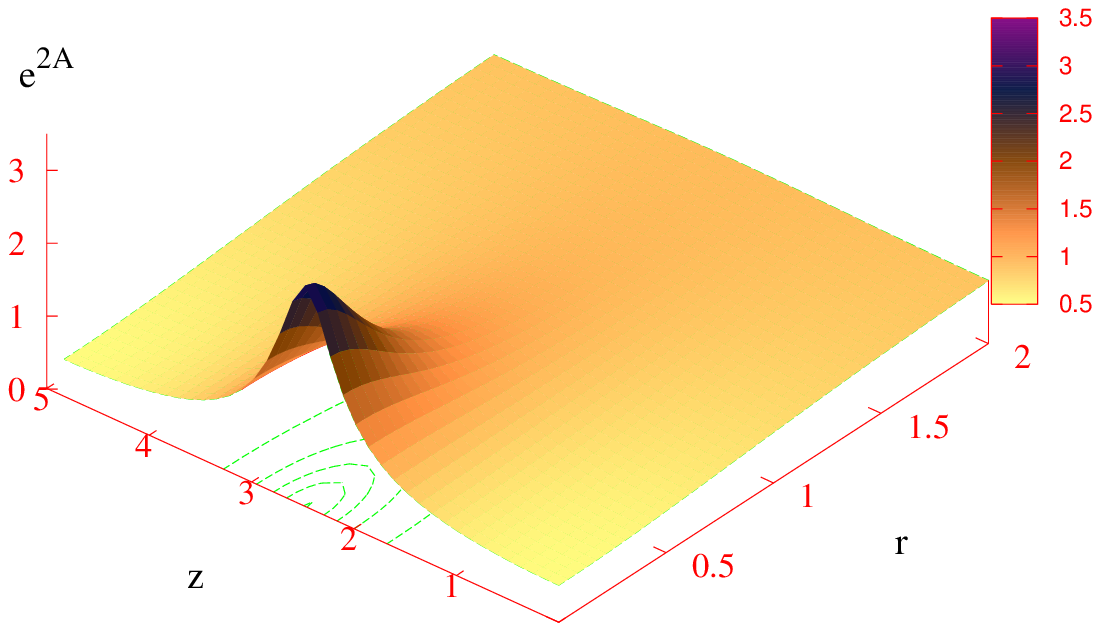,width=8cm}}
\put(7,12){\epsfig{file=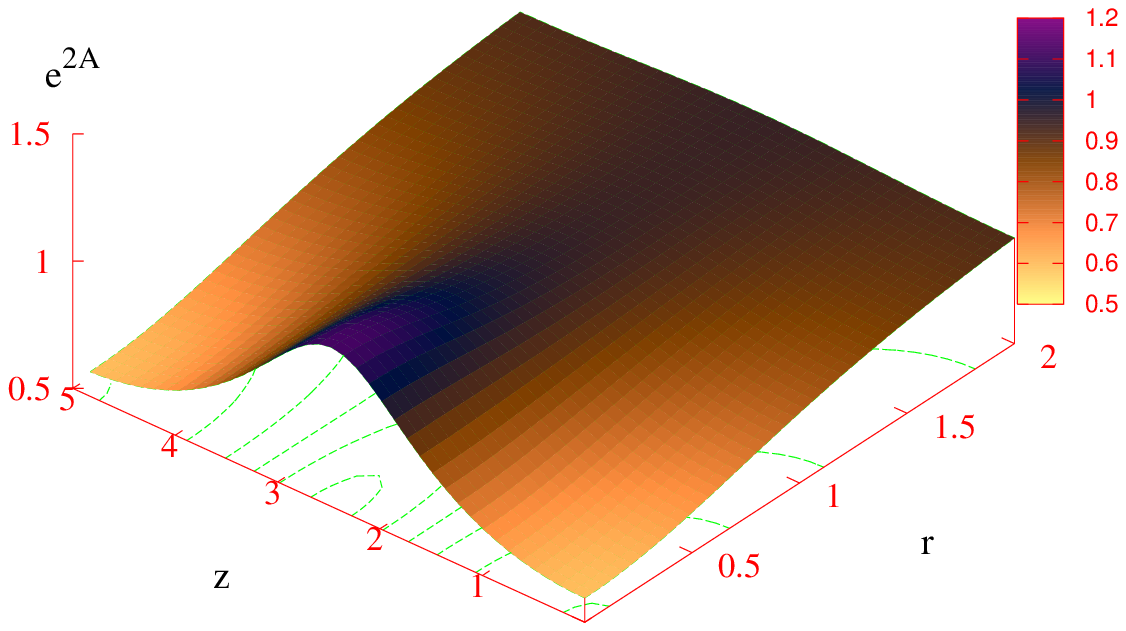,width=8cm}}
\end{picture}
\begin{figure}[h]
\caption{
The metric functions 
$e^{2A}$, $e^{2B}$, and $e^{2C}$ of a typical $d=6$ non-uniform string solution 
in Einstein-Gauss-Bonnet gravity 
are shown as functions of the   coordinates $r,z$
for $\alpha=0.036$ (left) and $\alpha=0.01$ (right). 
Both solutions have the same Hawking temperature $T_H=0.125$. }
\label{NUBS-3d}
\end{figure}

\begin{figure}[ht]
\hbox to\linewidth{\hss%
	\resizebox{8cm}{6cm}{\includegraphics{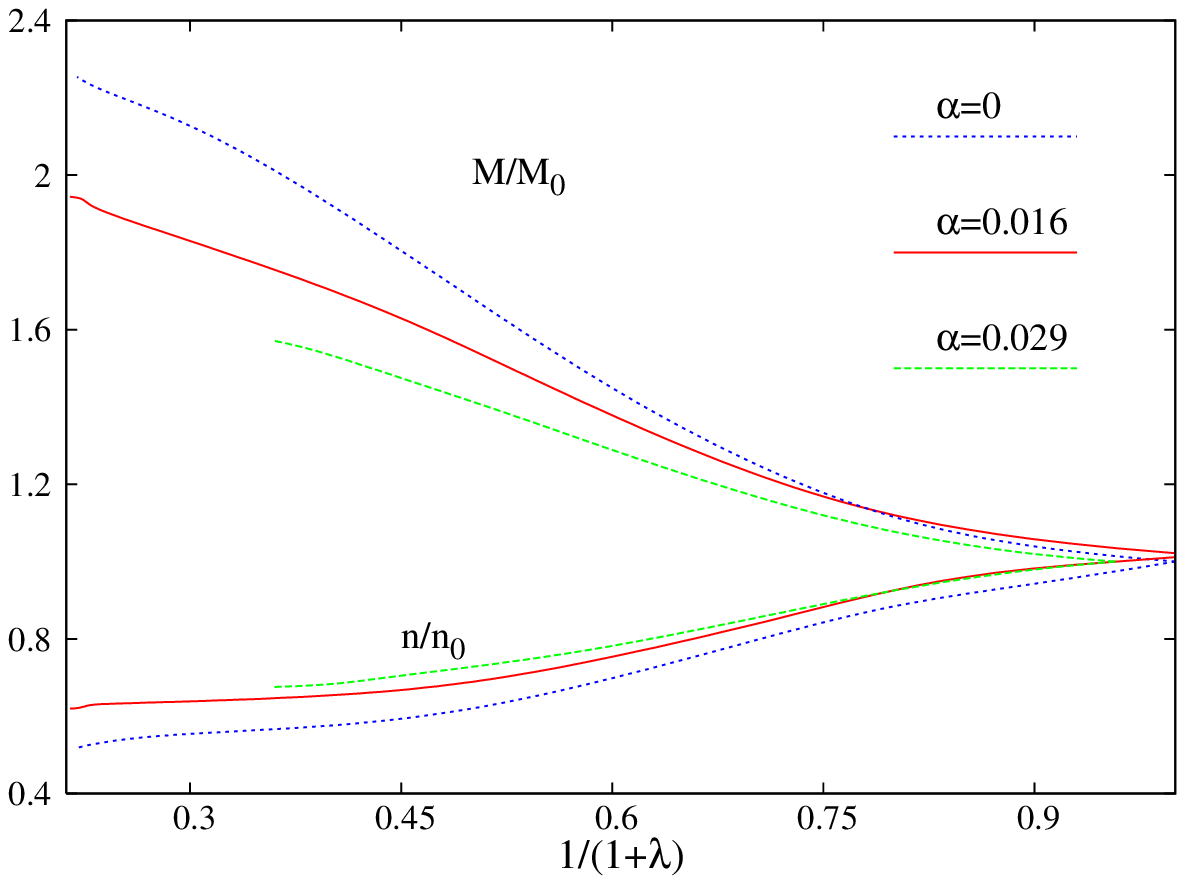}}
\hspace{5mm}%
        \resizebox{8cm}{6cm}{\includegraphics{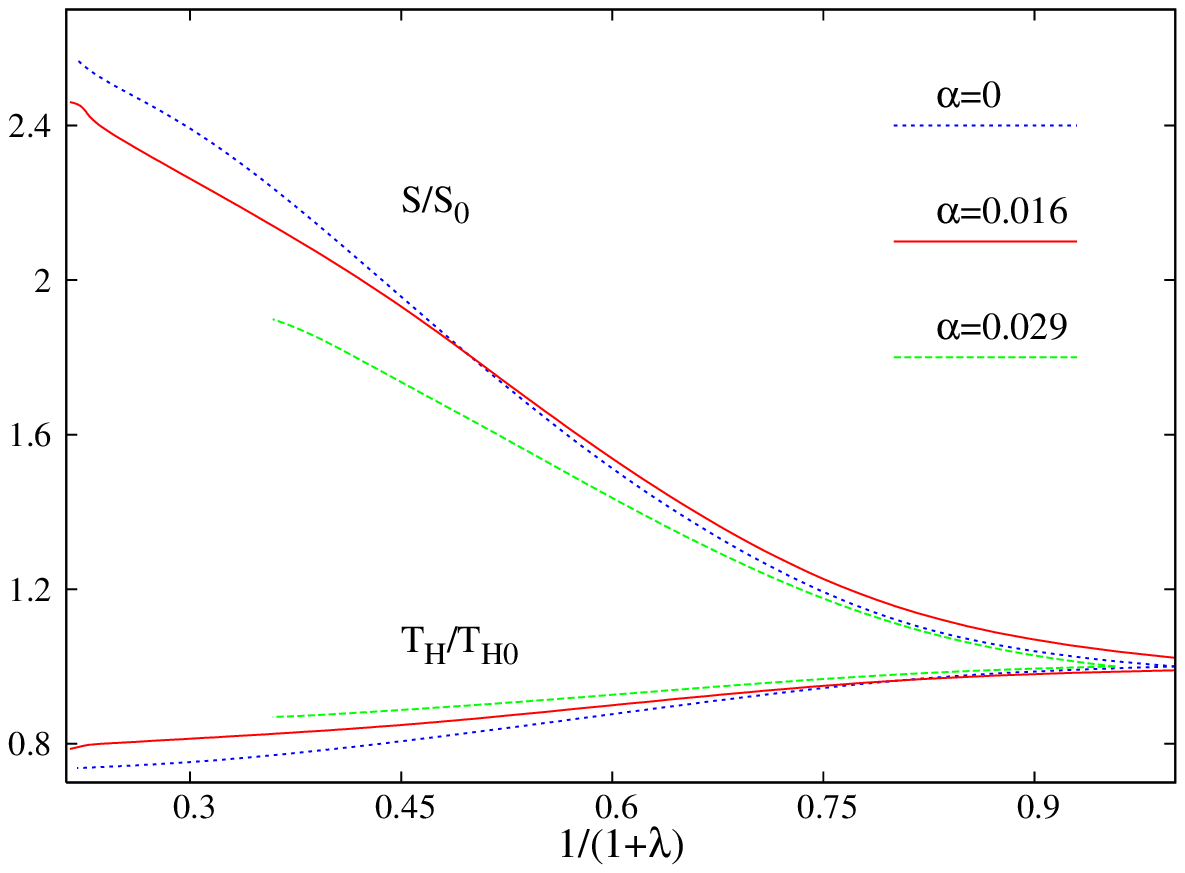}}	
\hss}

\caption{
{\small
The $M$, the relative tension $n$, the Hawking temperature $T_H$ and the entropy $S$ 
of the $d=6$ non-uniform string solutions
are shown 
 in units of the corresponding uniform string solution
(denoted by $M_0$, $n_0$, $T_0$, and $S_0$) as functions of
$1/(1+\lambda)$ (with $\lambda$
the non-uniformity parameter),  for several values of the Gauss-Bonnet coupling constant $\alpha$.}
}
\label{NUBS-lambda}
\end{figure}

\begin{figure}[ht]
\hbox to\linewidth{\hss%
	\resizebox{8cm}{6cm}{\includegraphics{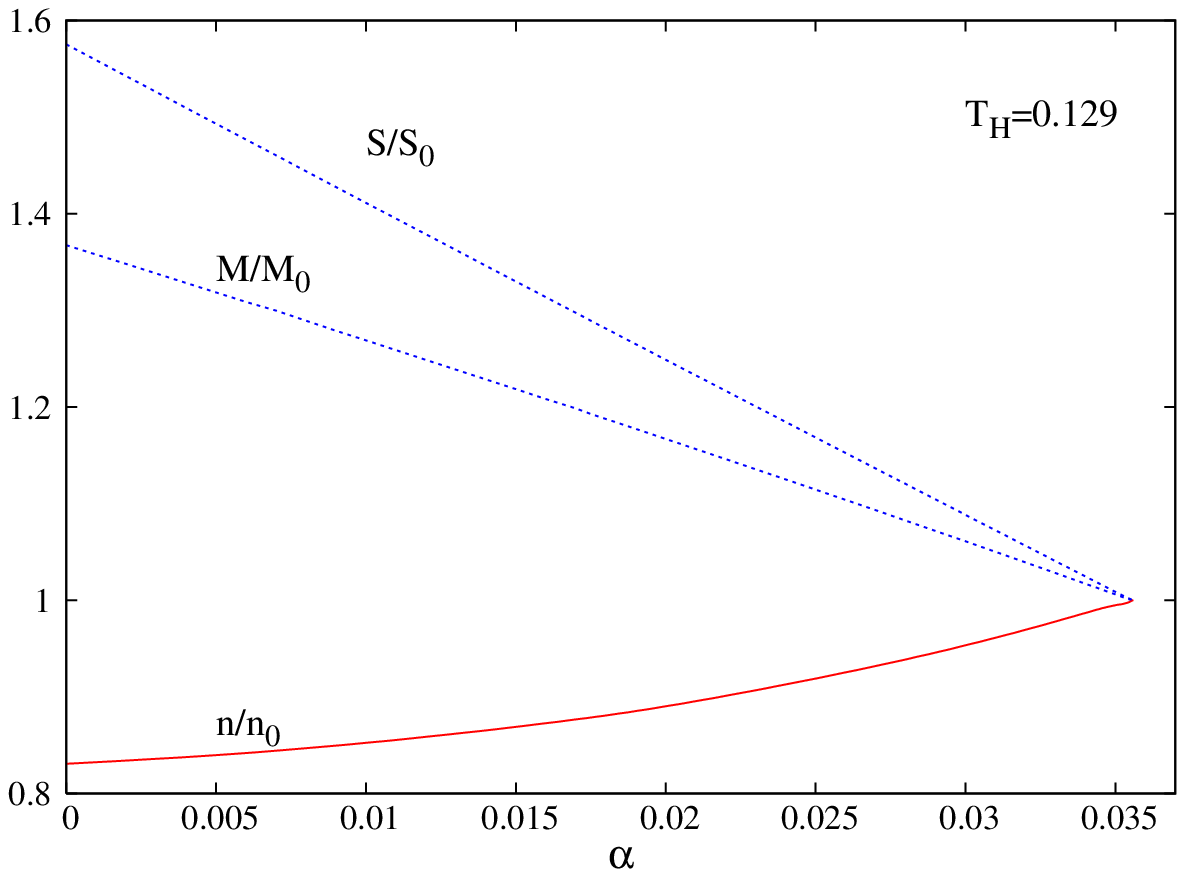}}
\hspace{5mm}%
        \resizebox{8cm}{6cm}{\includegraphics{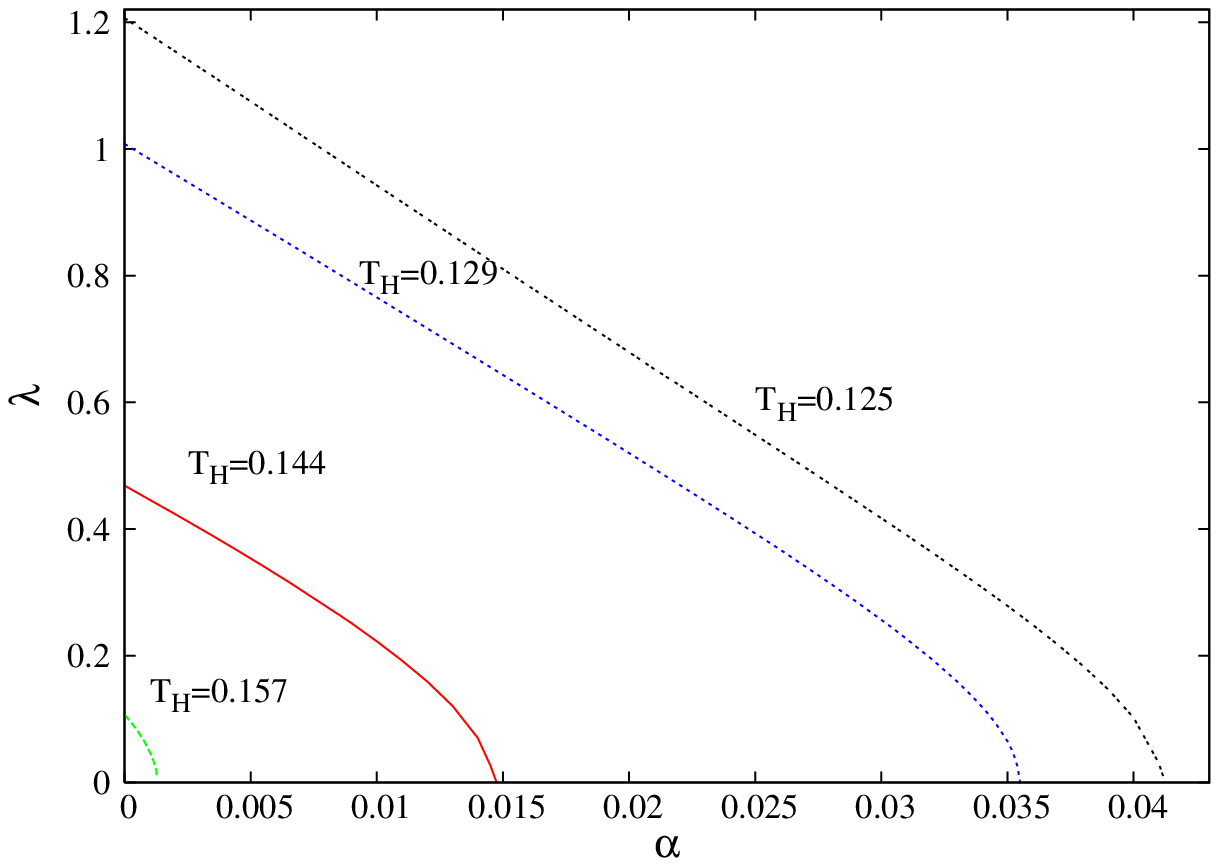}}	
\hss}

\caption{
{\small
{\it Left:} The $M$, the relative tension $n$   and the entropy $S$ 
of the $d=6$ non-uniform string solutions
are shown 
 in units of the corresponding uniform string solution
(denoted by $M_0$, $n_0$  and $S_0$) as functions of
the Gauss-Bonnet coupling constant $\alpha$  for a fixed value of the Hawking temperature.
{\it Right:} The deformation parameter $\lambda$ is shown as a function of the Gauss-Bonnet coupling constant $\alpha$  for several
 fixed values of the Hawking temperature.}
}
\label{NUBS-var-alpha}
\end{figure}

\newpage
Within our approach, we could find EGB configurations
with reasonable accuracy\footnote{Although solutions
seem to exist for larger $\lambda$,
the numerical accuracy decreases drastically in that case.
In particular, the constraint equations fail to be satisfied.} up to $\lambda \simeq 3.7$.

We exhibit in Figure \ref{NUBS-lambda} the mass $M$, the relative tension $n$,
the temperature $T_H$ and the entropy $S$ of the $d=6$  non-uniform string
solutions, in units of the corresponding uniform string solution,
versus the non-uniformity parameter $\lambda$, for several values of $\alpha$. 
For any value we have considered, the key feature here is that the relative entropy grows with $\lambda$,
such that the NUBS are more entropic 
than the uniform critical string.
Moreover, these solutions are also cooler than the critical GL strings.
This is the picture one finds  for $\alpha=0$.
Also, one can see that as $\alpha$ increases, the derivation of the various quantities from the corresponding
values of the uniform solution decreases.

Turning now to the dependence of the solutions on $\alpha$, we have constructed
 NUBSs  in EGB theory
starting with the $d=6$ Einstein gravity solutions in \cite{Kleihaus:2006ee}
and slowly increasing the value of the GB coupling constant.
Similar to the uniform string cases (Section 3) or to the black hole case (Appendix A), 
we have found numerical evidence that the known $d=6$
Einstein gravity NUBSs also admit EGB generalizations.
Keeping fixed the asymptotic length $L$ of the compact direction,
the horizon coordinate $r_0$ and the Hawking temperature $T_H$, 
the solutions change smoothly with GB coupling constant.
The results of the numerical integration in this case are shown in Figure \ref{NUBS-var-alpha}
for several values of the Hawking temperature (thus of the boundary parameter $d_0$).
One can see that as the value of $\alpha$ increases, 
the non-uniformity parameter $\lambda$ descreases and the critical 
uniform solution
with the given value of $T_H$ is approach.
Along this branch, the entropy and mass decrease, while the relative tension increases.
Therefore we observe again the tendency of the GB term
to reduce the non-uniformity
of solutions. 
It is tempting to link that to the occurrence of a negative
local `effective energy density', $\alpha H_{t}^t<0$, also for NUBS.
This would imply the existence of an  extra repealing force,
which would reduce the string tension and will
'smooth' the configurations.

\section{Conclusions}
The phase structure of the $d\geq 5$ Kaluza-Klein theory with a single $S^1$ direction
contains
three\footnote{However, at least for $d=5,6$, in addition to these phases, there are also
bubble-black hole sequences \cite{Emparan:2001wk}, which we do not consider here.} known phases of static solutions \cite{Obers:2008pj}.
These are the uniform black string phase, the non-uniform black string phase
and the localized black hole phase.
The main purpose of this work was to find how a Gauss-Bonnet term in the gravity action 
affects the properties of the first
two phases of solutions mentioned above,
with a $S^{d-3}\times S^1$ topology of the horizon.

In the UBS case, we have 
answered this question by 
constructing the solutions  using both analytical
and numerical methods, for $5\leq d\leq 10$.
Exact solutions were constructed by taking the
GB term as a perturbation from pure Einstein
gravity. The nonperturbative solutions were found by solving numerically the field equations.
We argue that the presence of a GB term in the langrangian leads to 
some interesting new features,
the cases $d=5,6$ being special.
For $d=5$ there is a mass gap in the spectrum of black strings.
 In six dimensions, 
there are 
thermodynamically stable UBS solutions for some range of the GB coupling constant.
Similar to the Einstein gravity case, the $d=5$ and $d>6$ black strings are thermodynamically unstable.

We have also discussed the issue of Gregory-Laflamme instability for the UBS solutions with 
a GB term.
All solutions were found to be unstable, except for the $d=6$ case.
There, for a given mass, the solutions with a large enough value of the GB coupling constant do not possess
a Gregory-Laflamme instability.
Thus
our results provide also an explicit realization of the Gubser-Mitra conjecture,
that correlates the dynamical and thermodynamical stability.

However, one may argue that since the action (\ref{action}) is a one loop string theory 
approximation, our results would be relevant for small value of $\alpha$ only. 
Therefore higher order corrections
may change this picture, rendering the $d=6$ solutions unstable as well.
Moreover, for $d>8$, higher order terms in the Lovelock hierarchy
would enter the gravity action.
Then we expect a more complicated picture in that case.

One should also remark that the uniform solutions in this work can also be
interpreted as black holes in a EGB-dilaton 
theory in $d-1$ dimensions.
The action of this model is found by doing a reduction  with respect to
the Killing vector $\partial/\partial z$, and has a very complicated
form, with non-standard kinetic terms for the dilaton  (see $e.g.$
the Appendix A in Ref. \cite{Bao:2007fx}).
(Note that the  asymptotically flat  five dimensional EGB-dilaton black holes will be thermally stable
for some range of the parameters.)
Following the usual procedure, one can add a U(1) charge to the  $d-1$ dimensional black hole solutions by boosting
a black string in the $(z,t)$-plane and performing a Kaluza-Klein reduction with respect to the 
new $z-$direction.
The properties of the resulting configurations are likely to be quite different as compared to the standard EGB-Maxwell 
black holes \cite{Torii:2005nh}.
We hope to return on these aspects in future work.

Concerning the NUBS branch of solutions,
we have reported some partial
results for the case $d=6$ only.
Our findings support the existence of EGB generalizations 
of the known Einstein gravity solutions. 
For the parameter range we have explored, we have notice a 
similar qualitative beaviour as in the absence 
of a GB term.
In particular, our results 
suggest that the black hole and the black string
branches merge at a topology changing transition
also in this case, at least for small enough values of the GB coupling constant.
There we have also noticed a tendency of the GB interaction to 
'smoothen' the configurations, decreasing the deviations from
the corresponding uniform solutions.

To push forward our understanding of these issues,  
it would be interesting to construct the corresponding caged black holes.
Given the results in this work, the dimensions $d=5,6$  are of particular interest,
and we expect the existence in these cases of caged black holes
 possessing a positive specific heat.

Also, the EGB black string solutions
should possess generalizations with matter fields,
the case of Einstein-Maxwell-dilaton theory being perhaps the most important.
However, the Harrison-type generation techniques used in  \cite{Kleihaus:2009ff}
to construct electrically 
charged solutions 
do not hold in the presence of a GB term, and
one is constrained to approach  this problem numerically.

\section*{Acknowledgements}
YB is grateful to the
Belgian FNRS for financial support.
 The work of ER was supported by a fellowship from the Alexander von Humboldt Foundation.
  
 \appendix 
\section{The Schwarzschild black hole in EGB theory}
The generalizations of the Schwarzschild black hole solutions in EGB gravity 
were first discovered by Boulware and  Deser \cite{Deser},
and Wheeler \cite{Wheeler:1985nh}, independently.
The general solutions are classified into the plus and minus branches.
In the $\alpha\to 0$ limit, the solutions in the 'minus branch' recover the ones
in the general relativity,
while there is no such limit in the 'plus branch'.
Restricting to the 'minus branch' solutions, the line element of the Schwarzschild black hole in EGB gravity  reads
 \begin{eqnarray}
\label{general-metric-form}
ds^2=\frac{dr^2}{f(r)}+r^2 d\Omega_{d-2}^2 -f(r) dt^2,
\end{eqnarray}
where
 \begin{eqnarray}
\label{N1}
&f(r)=1+\frac{2r^2}{\alpha(d-3)(d-4)}
 \bigg (
  1-\sqrt{1+ \frac{\alpha}{r_h^2}(d-3)(d-4)\left(1+(d-3)(d-4)\frac{\alpha}{4r_h^2}\right)\left(\frac{r_h}{r}\right)^{d-1}}
 ~\bigg) .~{~~~~}
\end{eqnarray}
In this relation, $r_h$ is the largest positive root of $f(r)$, typically associated to the outer horizon\footnote{The global structure
of this solutions is analyzed in \cite{Torii:2005xu}. Their generalizations with higher order curvature terms are discussed
$e.g.$ in \cite{cai1}.}
 of a black hole, $f(r_h)=0$. 
The EGB Schwarzschild solution exists for 
all $r_h>0$ and $\alpha \geq -2r_h^2/(d-3)(d-4)$.
As $\alpha\to 0$, one recovers to leading order the Schwarzschild black hole expression,
\begin{eqnarray}
\label{N4}
f(r)=1-(\frac{r_h}{r})^{d-3}-\frac{(d-3)(d-4)}{4r_h^2}(\frac{r_h}{r})^{d-3}\left(1-(\frac{r_h}{r})^{d-1}\right)\alpha+\dots
\end{eqnarray}
Also, as $r\to r_h$ one finds
\begin{eqnarray}
\label{N3}
f(r)=\frac{ (d-3)(4r_h^2+(d-4)(d-5)\alpha)}{4r_h^3+2(d-3)(d-4)r_h\alpha }(r-r_h)+O(r-r_h)^2.
\end{eqnarray}
Note that the usual form for $f(r)$ in the literature
is not given in terms of $r_h$ but rather in terms of a parameter $\mu=r_h^{d-5}(4r_h^2+(d-3)(d-4)\alpha)/4$. 
As seen bellow, this corresponds to fixing the mass of the solutions;
however, the expression (\ref{N1}) facilitates a comparison 
with the numerical uniform black string solutions in which case the control
parameter is the even horizon radius and not the mass, which is computed from the numerical output.

In terms of the dimensionless coupling constant $\beta=\alpha/r_h^2$,
the Hawking temperature of a Schwarzschild-Tangerlini solution in EGB theory is given by
 \begin{eqnarray}
\label{TH-BH}
T_H=\frac{(d-3)}{4\pi r_h}\frac{4 +(d-4)(d-5)\beta}{4 +2(d-3)(d-4)  \beta},
 \end{eqnarray}
 with 
  \begin{eqnarray}
\label{TH-BH0}
T_H=\frac{(d-3)}{4\pi r_h}\left(1-\frac{(d-1) (d-4) }{4 }\beta \right)+O(\beta)^2,
 \end{eqnarray}
in the small $\beta$ limit. Thus, for $\alpha>0$, the Hawking temperature of such systems appears to be suppressed relative
to that of a Einstein gravity black hole of equal horizon area.
 
At large distances ($r^2\gg \alpha$)
the EGB black hole behaves like the Schwarzschild solution.
Once the event horizon radius $r_h$ is fixed 
(as is done in the numerical construction of the solutions), 
the parameter $\alpha$ enters the leading order term in the asymptotics 
\begin{eqnarray}
\label{N2}
f(r)=1-\left(1+\frac{\alpha(d-3)(d-4)}{4r_h^2}\right)\left(\frac{r_h}{r}\right)^{d-3}+ \dots,
\end{eqnarray}
which fixes the value of the mass:
\begin{eqnarray}
\label{N21}
M=\frac{(d-2)V_{d-2}r_h^{d-3}}{16\pi G}\left(1+\frac{(d-3)(d-4)}{4 }\beta\right),
\end{eqnarray}
where $V_{d-2}$ is the area of the unit $S^{d-2}$ sphere.
One can see that the case $d=5$ is special, given the existence of a mass gap
$M> {3\pi} \alpha/ {16 G}$.
The entropy of the Schwarzschild black hole in  EGB theory, as computed from (\ref{S-Noether} ) is
\begin{eqnarray}
\label{BH-entropy}
 S= \frac{1}{4G}V_{d-2}r_h^{d-3}(1+ \frac{(d-2)(d-3)}{2}\beta).
\end{eqnarray}  
One can easily verify that the first law of thermodynamics 
$dM = T_H dS$ also holds.
 
This solution has a specific heat
\begin{eqnarray}
\label{heat}
C_p = T_H\left(\frac{\partial S}{\partial T_H}\right) 
= -\frac{(d-2)r_h^{d-2}(4 +(d-4)(d-5)\beta)(2 +(d-3)(d-4)\beta)^2 }
{8(8 +2(d-9)(d-4) \beta+(d-3)(d-4)^2(d-5)\beta^2)} .
\end{eqnarray}  
Without entering into details, 
we mention the existence of some substantial 
differences between the thermodynamics
of the $d=5$ EGB Schwarzschild solutions and their Einstein gravity counterparts (here we shall restrict to
the physical case $\alpha>0$).
If the black holes are large enough, $r_h\gg \sqrt{\alpha}$,
then they behave like their Einstein gravity counterparts since $C_p<0$.
A different picture is found for small values of $r_h$, 
with $T_H\simeq r_h$ in that case and thus $C_p>0$.
In fact, the specific heat changes its sign 
at  $r_h=\sqrt{\alpha}$.
This implies the existence of a branch 
of small five dimensional EGB black holes which is thermodynamically stable
(see the Ref. \cite{Garraffo:2008hu} for a review of these aspects).
 However, since $C_p<0$, the $d>5$ solutions are thermodynamically unstable for all values of $r_h,\alpha>0$. 
 
The classical stability of the Schwarzschild black hole in EGB theory  has been discussed in \cite{Konoplya:2008ix} for $5\leq d\leq 11$.
 The results there show that an instability occurs for $d=5,6$ 
at some large values of $\alpha$ (see also \cite{hiraya}).    
   

\end{document}